\documentclass[3p,12p]{elsarticle}
\usepackage{lineno,hyperref}
\usepackage{comment}
\usepackage{epsfig}
\usepackage{epstopdf}
\usepackage{siunitx}
\usepackage{amsmath}
\usepackage{graphicx}
\usepackage{caption}
\usepackage{subcaption}
\modulolinenumbers[1]
\journal{Journal of the Mechanics and Physics of Solids}
\makeatletter
\def\ps@pprintTitle{%
	\let\@oddhead\@empty
	\let\@evenhead\@empty
	\def\@oddfoot{\centerline{\thepage}}%
	\let\@evenfoot\@oddfoot}
\makeatother
\begin{document}
	\begin{frontmatter}
		
		\title{Buckling initiation in layered hydrogels during transient swelling}
		
		\author[mymainaddress,mysecondaddress]{Arne Ilseng \corref{mycorrespondingauthor}}
		\cortext[mycorrespondingauthor]{arne.ilseng@ntnu.no}
		\author[mymainaddress]{Victorien Prot}
		\author[mymainaddress]{Bj\o rn H. Skallerud}
		\author[mysecondaddress]{Bj\o rn T. Stokke}
		\address[mymainaddress]{Biomechanics, Department of Structural Engineering, NTNU, Norwegian University of Science and Technology, 7491 Trondheim, Norway}
		\address[mysecondaddress]{Biophysics and Medical Technology, Department of Physics, NTNU, Norwegian University of Science and Technology, 7491 Trondheim, Norway}

\begin{abstract}
Subjected to compressive stresses, soft polymers with stiffness gradients can display various buckling patterns. These compressive stresses can have different origins, like mechanical forces, temperature changes, or, for hydrogel materials, osmotic swelling. Here, we focus on the influence of the transient nature of osmotic swelling on the initiation of buckling in confined layered hydrogel structures. A constitutive model for transient hydrogel swelling is outlined and implemented as a user-subroutine for the commercial finite element software Abaqus. The finite element procedure is benchmarked against linear perturbation analysis results for equilibrium swelling showing excellent correspondence. Based on the finite element results we conclude that the initiation of buckling in a two-layered hydrogel structure is highly affected by transient swelling effects, with instability emerging at lower swelling ratios and later in time with a lower diffusion coefficient. In addition, for hard-on-soft systems the wavelength of the buckling pattern is found to decrease as the diffusivity of the material is reduced for gels with a relatively low stiffness gradient between the substrate and the upper film. This study highlights the difference between equilibrium and transient swelling when it comes to the onset of instability in hydrogels, which is believed to be of importance as a fundamental aspect of swelling as well as providing input to guiding principles in the design of specific hydrogel systems.
\end{abstract}


\end{frontmatter}

%\linenumbers

%%%MAIN TEXT%%%%
\section{Introduction}
\begin{figure}[h]
 \centering
 \includegraphics[width=8.5cm]{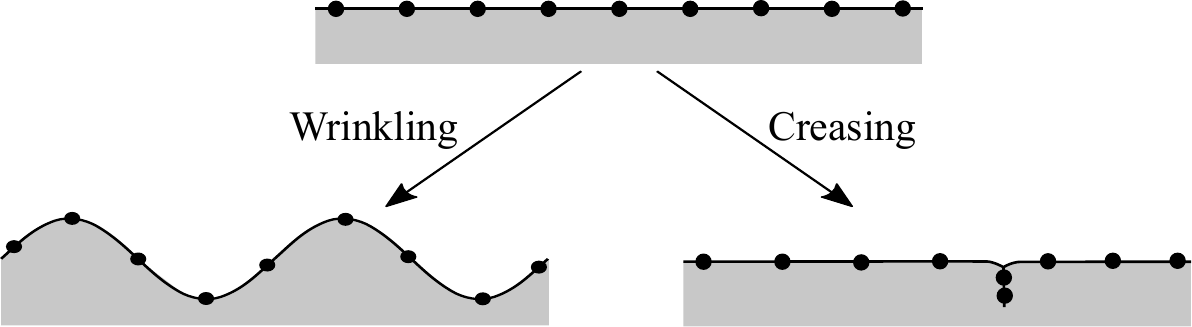}
 \caption{Illustration of wrinkling and creasing in the surface of a swelling gel.}
 \label{f:wrink_crease}
\end{figure}

Hydrogels are polymer networks swollen in an aqueous solution. Natural entities, e.g. extracellular matrix or the vitreous body of the eye, can be understood within this concept. Similar, various synthetic hydrogels are exploited in commercial products due to their particular properties (e.g. disposable diapers and contact lenses), and have a significant potential for applications like smart valves \cite{Beebe2000}, tissue engineering \cite{Chan2008,Annabi2014,Seliktar2012}, drug delivery systems \cite{Bysell2006, Li2016a, Culver2017}, and biological sensors \cite{Tierney2008,Buenger2012}. For many of these applications, the capability of hydrogels to swell or shrink as a response to stimuli (e.g. changes in temperature, mechanical forces, pH, salinity level, electric field, specific molecules recognized by included capture moieties) is exploited. During this swelling process, instabilities can occur at the surface of the gels causing wrinkling (i.e. global harmonic waves) and/or creasing (i.e. localized sharp folds) \cite{Tanaka1987,Chan2006,Trujillo2008,Chan2008a,Guvendiren2009,Breid2009,Chung2009,Yang2010,Li2012a,Chen2014,Zhou2017}, as illustrated in Figure \ref{f:wrink_crease}. While creasing mainly occurs as the first mode of instability when homogeneous gels are exposed to large swelling \cite{Trujillo2008, Li2012a}, wrinkling patterns are first and foremost found in gels having a gradient stiffness through the thickness, caused by a variation in the crosslinking density within the gel \cite{Guvendiren2009,Guvendiren2010,Guvendiren2010a}, or by the deposition of a thin and stiff film at the outer surface of the gel, possibly to alter properties like permeability, stability or biocompatibility \cite{Sultan2008, Prot2013, Sherstova2016}. In addition, the impact of the mechanically layered or anisotropic character of natural entities on morphogenesis attains growing interest \cite{Liu2012,Li2012a} with structure formation in aging skin as an example \cite{Limbert2018}. 

A phenomenon closely related to swelling induced instability in layered hydrogels is the evolution of buckling patterns in layered polymer plates under mechanical loading \cite{Bowden1998,Groenewold2001,Yang2010,Yin2018}. Assuming linear elastic material behavior, analytical expressions can be found for the critical strain at the onset of buckling and the wavelength of the resulting wrinkling pattern based on the stiffness ratio between the film and the substrate and the thickness of the film \cite{Stafford2004}. Along the same lines, linear perturbation analysis (LPA) frameworks have been developed for studying the onset of instability in layered hydrogel systems exposed to equilibrium swelling \cite{Wu2013,Wu2017}, i.e. assuming a homogeneous field for the chemical potential in the gel at all times. While these analytical or semi-analytical approaches are computationally efficient, they neglect the transient nature of the osmotic swelling process in hydrogels, giving a time-dependent and inhomogeneous field of the chemical potential through the gel, which could possibly alter buckling initiation. Accounting for the effects of transient swelling in layered gels calls upon numerical procedures like the finite element method (FEM). 

For finite element simulations of the equilibrium swelling of gels, numerous constitutive models are available in the literature, both general formulations defining the gel by its basic properties \cite{Hong2009,Kang2010a} and formulations that relate the parameters in the simulations more explicitly to the network properties of the specific gels \cite{Marcombe2010,Prot2013}. To also describe the transient nature of hydrogel swelling there are generally two methods that are used in the literature. The most available approach is to utilize the analogy between gel diffusion and heat transfer and implement the swelling process using the thermal modeling capabilities already available in commercial finite element codes \cite{Toh2013,Duan2013}. The other and more general approach is to formulate diffusion-deformation specific elements \cite{Zhang2009,Chester2015,Bouklas2015}, however, the implementation of such elements can be considered challenging and time-consuming.

The studies in the literature using finite element simulations for predicting the onset of instability mainly focus on buckling caused by equilibrium swelling (i.e. omitting transient effects) using 2D \cite{Kang2010,Wu2013,Weiss2013,Budday2015} or 3D models \cite{Xu2014a,Xu2015,Xu2016}. Transient effects related to swelling induced buckling are less described. Bouklas \textit{et al.} \cite{Bouklas2015} briefly discussed simulations of transient swelling, however, they did not demonstrate the difference between equilibrium and transient swelling for the initiation of buckling. Toh \textit{et al.} \cite{Toh2016} only considered plates with homogeneous material properties, in addition, they used a multi-point constraint routine to force a sinusoidal buckling pattern and hence neglected creasing as a possible mode of instability. In their recent study, Yu \textit{et al.} \cite{Yu2018} investigated buckling initiation in a 2D model of a thin constrained plate exposed to transient swelling, however, they did not consider gels with a stiffness gradient. Finally, Dortdivanlioglu and Linder \cite{Dortdivanlioglu2019} studied the initiation of buckling in layered confined hydrogels during transient swelling, though, they did not discuss the dependence of the diffusion coefficient for the initiation of instability nor the effective swelling ratio at the onset of buckling. In addition, none of the mentioned studies discuss the onset of instability during transient swelling of soft-on-hard layered gels. 

Although buckling caused by mechanical compression or equilibrium swelling of soft layered materials is well studied using experimental, analytical, or numerical procedures, the effect of the transient nature of hydrogel swelling for the onset of buckling remains insufficiently understood. Hence, we address the effects of transient swelling on the critical swelling ratio, the time to buckling, and the wavelength of the resulting buckling pattern for confined hydrogels. Furthermore, we study both hard-on-soft systems, with stiffness ratios between the film and the substrate in the moderate (20, 100) and low range (2, 5), and a soft-on-hard system with a stiffness ratio of 0.5. Due to its importance for the presented simulation results, we give a detailed discussion on the introduction of surface imperfections, the quantification of onset of instability, and the use of a plane strain assumption. For the case of equilibrium swelling, we benchmark our model against an LPA framework available in the literature \cite{Wu2013}, obtaining excellent correspondence. The results of this study are believed to be of importance for the development of models that can predict hydrogel swelling behavior in a general manner, as well as providing valuable insight into the nature of swelling induced buckling processes in layered hydrogels, which can affect guiding principles for the design of hydrogel systems.

The article is organized as follows: In the next section, the geometrical properties of the problem studied herein are described. In Section \ref{sec:mod_form}, the constitutive formulation used for transient hydrogel swelling is presented. Section \ref{sec:LPA} briefly describes the LPA approach used to predict the onset of buckling during equilibrium swelling as a benchmark for the finite element approach. Thereafter, in Section \ref{sec:fe_mod}, the finite element modeling is described in detail, including the implementation of the constitutive model for the commercial finite element software Abaqus and a validation case. Section \ref{sec:res} presents and discusses the results obtained from the finite element procedure and compares them to the linear perturbation results for the case of homogeneous swelling. Finally, conclusions from the study are presented in Section \ref{sec:conc}. 

\section{Problem definition}\label{sec:prob_def}
\begin{figure}[h]
\centering
  \includegraphics[width=8.5cm]{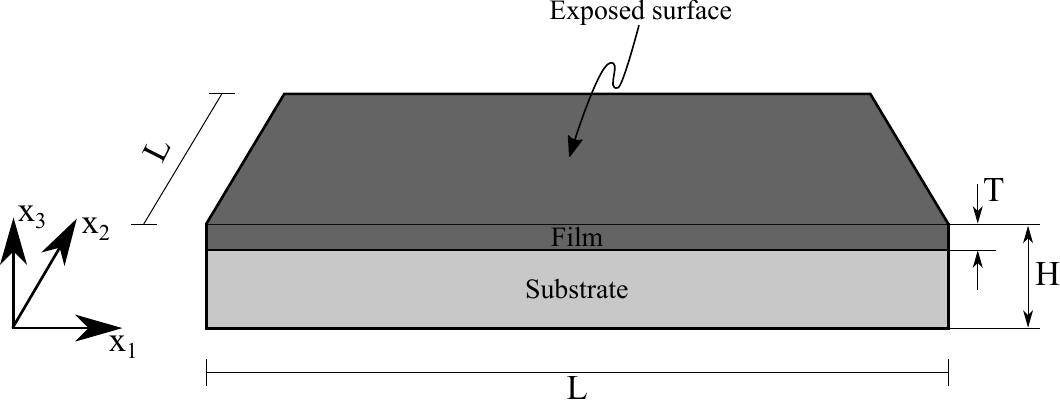}
  \caption{Illustration of a layered hydrogel plate in the coordinate system $\mathbf{x}$ where only the upper free surface is exposed to a solvent.}
  \label{f:plate}
\end{figure}

To study the onset of swelling induced instability in constrained layered hydrogel plates we define a square plate as illustrated in Figure \ref{f:plate}, having edges of length $L$ and a total thickness of $H$. The upper part of the plate consists of a film with a contrast stiffness and a thickness $T$. In the present study, the lengths of the plate are set to $L=8$ mm, the total plate thickness is set to $H=0.5$ mm, while the thickness of the film is set to $T=$\SI{50} {\micro \meter}. We introduce the film to plate thickness ratio $\eta=T/H$, and note that a value of $\eta=0.1$ is used herein. The influence of the chosen dimensions on the obtained results is discussed in Section \ref{sec:res_inidim}. 

All lateral edges of the plate are constrained from in-plane deformation, while there are no restrictions to swelling in the out-of-plane direction. At its upper surface, the plate is exposed to a solvent. Hence, a change in the chemical potential in the surrounding of the gel will cause an inhomogeneous and time-dependent profile for the chemical potential through the thickness of the gel. This chemical potential can cause swelling and hence give the plate a new and time-dependent total height of $h$. The chemical potential profile through the thickness of the gel will depend on the diffusion coefficient of the solvent molecules. As the goal of this study is to investigate the effect of transient swelling on the onset of buckling, we study the described problem with various values for the diffusion coefficient of the solvent molecules. Calculations are performed for diffusion coefficients between $10^{-11}$ m$^2$/s and $1$ m$^2$/s. While $10^{-11}$ m$^2$/s represents the lower range of physical values for a gel swelling in water \cite{Caccavo2018}, $1$ m$^2$/s is an unrealistic value, however, it is used to obtain buckling results when the chemical potential is approximately homogeneous through the gel. 

\section{Constitutive formulation}\label{sec:mod_form}
\subsection{Equilibrium swelling} \label{sec:mat_mod}
The constitutive modeling of hydrogel behavior applied herein is based on the work by Hong \textit{et al.} \cite{Hong2008,Hong2009}, Kang and Huang \cite{Kang2010a}, and Toh \textit{et al.} \cite{Toh2013}. The free-energy function for the hydrogel is assumed to originate from the additive contributions of stretching of the polymer network and mixing of the polymer and the solvent molecules \citep{flory1953,Huggins1941,Flory1942} 
\begin{linenomath}
\begin{equation}
\begin{split}
W\left(\mathbf{F},C\right)= &\frac{1}{2}NkT\left(I_1-3-2\ln J\right) +  \\ & \frac{kT}{v}\left(vC\ln\left(\frac{vC}{1+vC}\right)+\frac{\chi vC}{1+vC}\right)
\label{eq:pot_func}
\end{split}
\end{equation}
\end{linenomath}
where $N$ is the number of polymeric chains per reference volume, $kT$ is the temperature in the unit of energy, $\mathbf{F}$ is the deformation gradient tensor, $I_1=\text{tr}\mathbf{b}$ is the first invariant of the left Cauchy-Green tensor $\mathbf{b}=\mathbf{F}\mathbf{F}^\text{T}$,  $J=\det{\mathbf{F}}$ is the volume ratio of the material, $v$ is the volume per solvent molecule, $C$ is the nominal concentration of solvent molecules, and $\chi$ is the Flory-Huggins parameter. 

By assuming that both the polymer network and the solvent molecules retain their volumes through the swelling process, we find that the volume increase of the gel only can come from an increase in the number of solvent molecules inside the gel, hence we can write 
\begin{linenomath}
\begin{equation}
J = \det{\mathbf{F}} = 1+vC
\label{eq:mol_incom}
\end{equation}
\end{linenomath}
Further, to enable implementation for the finite element method, we can introduce a new free-energy function $\hat{W}$ by the use of a Legendre transformation as 
\begin{linenomath}
\begin{equation}
\hat{W}\left(\mathbf{F},\mu\right)=W\left(\mathbf{F},C\right)-\frac{\mu}{v}\left(J-1\right)
\label{eq:pot_func_hat}
\end{equation}
\end{linenomath}
and hence ensure the deformation gradient $\mathbf{F}$ and the chemical potential $\mu$ to be the two independent variables of the model. The Cauchy stress tensor $\boldsymbol{\sigma}$ can then be obtained from the free-energy potential function as
\begin{linenomath}
\begin{equation}
\boldsymbol{\sigma}=\frac{1}{J}\frac{\partial \hat{W}\left(\mathbf{F},\mu\right)}{\partial \mathbf{F}}\mathbf{F}^\text{T} = NkT\left(J^{-1}\mathbf{b}+\frac{1}{Nv}\left(\ln\frac{J-1}{J}+\frac{1-Nv}{J}+\frac{\chi}{J^2}-\frac{\mu}{kT}\right)\mathbf{I}\right)
\label{eq:cauchy_stress}
\end{equation}
\end{linenomath}
where $\mathbf{I}$ is the second-order identity tensor.

Normalized quantities are applied in the following and are defined according to $\bar{W}=\hat{W}/NkT$, $\bar{\boldsymbol{\sigma}}=\boldsymbol{\sigma}v/kT$ , and $\bar{\mu}=\mu/kT$.

\subsection{Kinetics}
In order to capture transient swelling in the constitutive model, the migration of solvent molecules into the hydrogel network must be accounted for. Here, we adopt the modeling procedure by Hong \textit{et al.} \cite{Hong2008} and assume that the small solvent molecules diffuse in the hydrogel network. A kinetic law for the gel can be written as 
\begin{linenomath}
\begin{equation}
\mathbf{\Phi} = -\mathbf{M}\frac{\partial \mu}{\partial \mathbf{X}}
\label{eq:JK}
\end{equation}
\end{linenomath}
where $\boldsymbol{\Phi}\left(\mathbf{X},t\right)$ is the nominal flux vector for the solvent molecules, $\mathbf{X}$ denotes the position of the material points in the reference configuration, while $\mathbf{M}$ is the mobility tensor. 

Further, the diffusion coefficient of the solvent molecules $D$ is taken as isotropic and independent of the deformation gradient $\textbf{F}$ and the nominal concentration of solvent molecules $C$ \cite{Hong2008}. The true flux vector $\boldsymbol{\phi}(\boldsymbol{x},t)$ is assumed to follow Fick's first law \cite{Feynman1963}
\begin{linenomath}
\begin{equation}
\boldsymbol{\phi}=-\frac{1}{J}\frac{CD}{kT}\frac{\partial\mu}{\partial \mathbf{x}}=-\frac{CD}{J}\frac{\partial\bar{\mu}}{\partial \mathbf{x}}
\label{eq:jk}
\end{equation}
\end{linenomath}
where $\mathbf{x}=\mathbf{x}\left(\mathbf{X},t\right)$ is the position of the material points in the current configuration. The nominal and true flux vectors can be related through
\begin{linenomath}
\begin{equation}
\boldsymbol{\phi}=\frac{\mathbf{F}}{J}\boldsymbol{\Phi}
\label{eq:jiJK}
\end{equation}
\end{linenomath}
Combining Equations (\ref{eq:mol_incom}), (\ref{eq:JK}), (\ref{eq:jk}), and (\ref{eq:jiJK}) we find that the mobility tensor can be written as
\begin{linenomath}
\begin{equation}
\mathbf{M}=\frac{D}{vkT}\mathbf{F}^\text{-1}\mathbf{F}^\text{-T}\left(J-1\right)
\label{eq:MKL}
\end{equation}
\end{linenomath}

It can be noted that the outlined material model neglects possible viscoelastic effects in the polymer network. The benefit of this approach is that the transient swelling is the only time-dependent feature of the model, and it is thereby easy to isolate the transient effects. The impact of this simplification on the obtained results is discussed in Section \ref{sec:time_initiation}. 

\section{Linear perturbation analysis}\label{sec:LPA}
Based on the work by Wu \textit{et al.} \cite{Wu2013} for a bilayer hydrogel we implemented an LPA framework for studying the onset of mechanical instability during equilibrium swelling (i.e. assuming a homogeneous chemical potential). The procedure is based on a 2D perturbation analysis where the perturbed deformation gradient $\tilde{\mathbf{F}}$ is assumed to take the form 
\begin{linenomath}
\begin{equation}
\tilde{\mathbf{F}} = 
\begin{bmatrix}
1+\frac{\partial u_1}{\partial x_1} & 0 & \lambda \frac{\partial u_1}{\partial x_3} \\
0 & 1 & 0 \\ 
\frac{\partial u_3}{\partial x_1} & 0 & \lambda\left(1+\frac{\partial u_3}{\partial x_3}\right) 
\end{bmatrix}
\end{equation}
\end{linenomath}
letting $\lambda$ be the out-of-plane deformation, while the assumed perturbations from the equilibrium state in the $x_1$- and $x_3$-direction are given by $u_1=U_1\left(x_3\right)\sin\left(\omega x_1\right)$ and $u_3=U_3\left(x_3\right)\cos\left(\omega x_1\right)$. The further steps of the method are thoroughly described in Wu \textit{et al.} \cite{Wu2013} and its implementation culminates in solving the determinant of an 8$\times$8 matrix equal to zero, where all nonzero elements are given between Equations (43) and (44) in Wu \textit{et al.} \cite{Wu2013}. 

\begin{figure}[h]
	\centering
	\includegraphics[width=10cm]{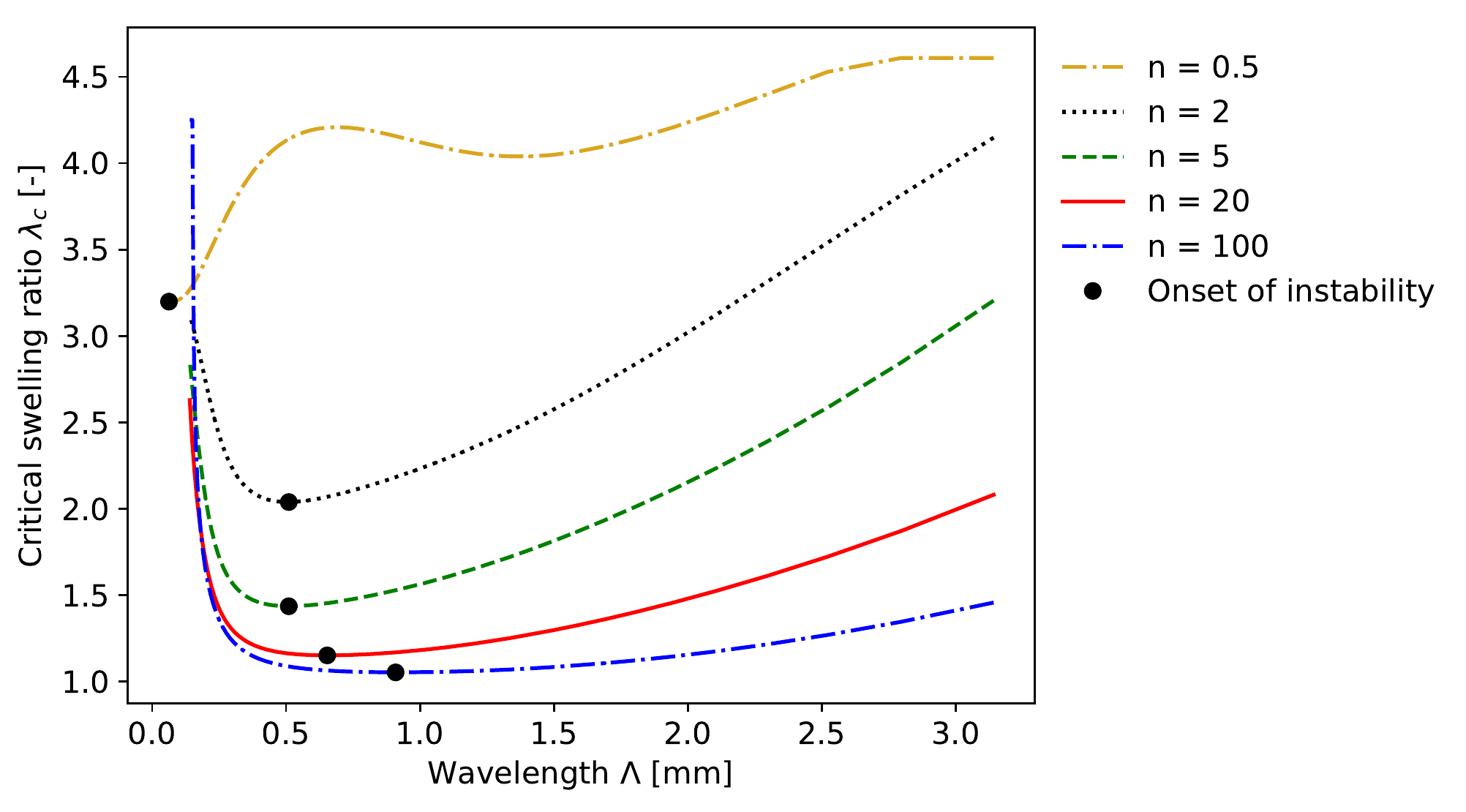}
	\caption{Critical swelling ratios as a function of the normalized perturbation wave number for $Nv_s=0.001$. The black markers at the minimum points in the diagram indicate the predicted critical swelling ratio and wavelength for the initiation of buckling.}
	\label{f:LPA}
\end{figure}

A stability plot, showing the estimated critical swelling ratio at the onset of instability for different wavelengths of the initial harmonic perturbation, is given in Figure \ref{f:LPA}. Here, $n$ denotes the stiffness ratio between the film and the substrate, $n=Nv_f/Nv_s$, with $Nv_f$ and $Nv_s$ being the stiffness of the film and the substrate, respectively. For readability, only results with the stiffness of the substrate set to $Nv_s=0.001$ is shown, however, similar trends would be found for $Nv_s=0.01$. The analysis is based on the plate height and film thickness as given in Section \ref{sec:prob_def}, while the plate width is assumed infinite in the LPA framework. 

For the soft-on-hard system ($n=0.5$) we can see that the minimum point for the critical swelling ratio occurs at the zero-wavelength limit. This indicates creasing as the dominating mode of instability and a similar response was found for soft-on-hard systems with stiffness ratio values of 0.1 and 0.8 (results not shown here due to readability). For the hard-on-soft systems ($n>1$) on the other hand, the minimum point of the critical swelling ratio is found at a defined wavelength, indicating wrinkling to be the dominating mode of instability. This difference between soft-on-hard and hard-on-soft systems is in accordance with previously reported results \cite{Wu2013}. 

From Figure \ref{f:LPA} it can also be noted that for hard-on-soft systems the critical swelling ratio increases as the stiffness ratio is reduced. Further, the instability curves for the hard-on-soft gels display a wider plateau in buckling wavelength around the critical point as the stiffness ratio is increased. Hence, it can be expected that the critical wavelength obtained in experiments would be more sensitive to experimental uncertainties, like the initial surface geometry, as the stiffness ratio is increased. Along the same lines, we can expect that the critical wavelength obtained using numerical methods as FEM could be more sensitive to numerical approximations as the stiffness ratio is increased. 

To study the effect of transient swelling for the initiation of buckling, we will make use of the finite element method as described in the following section. The swelling ratio and the wavelength obtained by the LPA will be used as a benchmark for the finite element simulations with a large diffusion coefficient (i.e.  $1$ m$^2$/s), causing a nearly homogeneous chemical potential through the thickness of the gel, to mimic the assumptions of the LPA calculations.

\section{FE modeling}\label{sec:fe_mod}

\subsection{Implementation and validation of the constitutive model}
\label{sec:mat_mod_imp}
\subsubsection{Implementation}
\begin{figure}[h]
\centering
  \includegraphics[width=9cm]{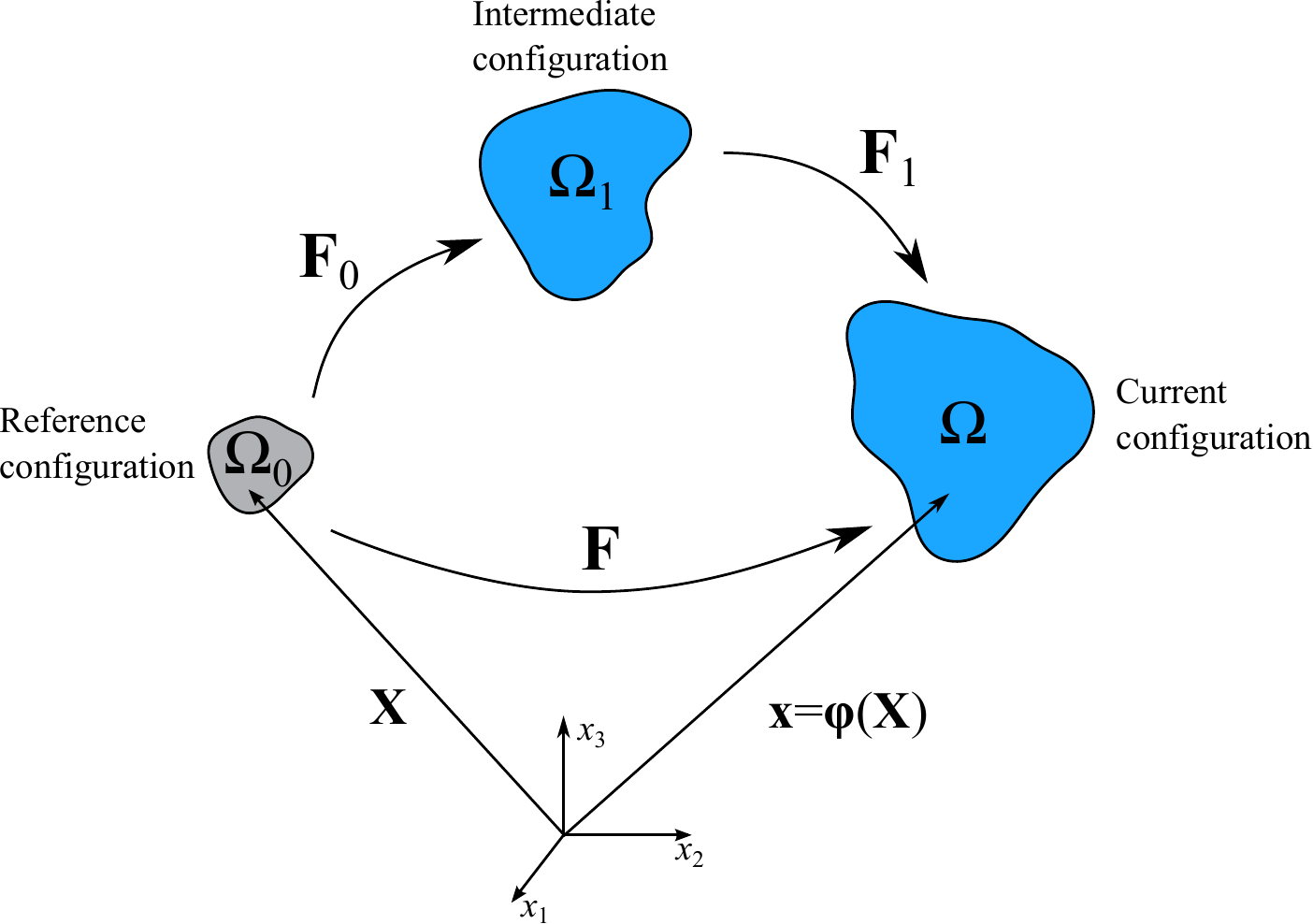}
  \caption{Illustration of the reference, intermediate, and current configurations. All finite element simulations herein start in the intermediate configuration.}
  \label{f:conf_split}
\end{figure}

To accommodate for finite element simulations of the defined swelling problem, the constitutive model described in Section \ref{sec:mat_mod} was implemented as a Fortran routine to be used with the commercial finite element software Abaqus/Standard. The equilibrium behavior of the model is implemented as a user-defined material, UMAT \cite{Abaqus2014}, defining the equations for the normalized stress $\bar{\boldsymbol{\sigma}}$ and the Jacobian tangent stiffness. As the chemical potential is singular in the dry state of the hydrogel $\Omega_0$, an intermediate configuration $\Omega_1$ is introduced. Consequentially, the full deformation tensor $\mathbf{F}$ can be split in a multiplicative manner as $\mathbf{F}=\mathbf{F}_1\mathbf{F}_0$. $\mathbf{F}_0$ maps the reference configuration to the intermediate configuration, and $\mathbf{F}_1$ maps the intermediate configuration to the current configuration, as illustrated in Figure \ref{f:conf_split}. In the intermediate configuration, the gel is assumed stress-free and in a state of isotropic strain such that $\mathbf{F}_0=\lambda_0\mathbf{I}$. In the present study, all finite element simulations start in the intermediate configuration $\Omega_1$ with a homogeneous distribution of the normalized chemical potential $\bar{\mu}_1$ in the gel and dimensions as given in Section \ref{sec:prob_def}. The value of $\lambda_0$ is found by numerically solving $\bar{\boldsymbol{\sigma}}=\mathbf{0}$ at the start of the finite element simulation for the initial chemical potential and the material parameters of the hydrogel. 

To account for the transient behavior of the gel swelling process, a coupled solvent diffusion and large deformation procedure is utilized. This procedure follows the main ideas presented by Toh \textit{et al.} \cite{Toh2013} and a detailed description is included in \ref{app:heat_ana}. 

The implemented Fortran code and an example input file for Abaqus is made available as a Mendeley dataset linked to this work \cite{Ilseng}.

\subsubsection{Validation}
To validate the implemented model, we study the case of 1D swelling of a homogeneous perfect plate. We start from an intermediate state $\Omega_1$ where the gel is in a state of isotropic swelling $\left(\mathbf{F}_0=\lambda_0\mathbf{I}\right)$ so that it equilibrates at a homogeneous chemical potential of $\bar{\mu}_1 = -2$. The plate is then confined from in-plane expansion but can still freely swell out-of-plane. The chemical potential at the surface of the plate is changed to $\bar{\mu}\left(X_3=H, t\right)=0$ and the gel gradually swells in the out-of-plane direction giving a deformation gradient with the non-zero elements $F_{11}=F_{22}=\lambda_0$ and $F_{33}=\lambda\lambda_0$ where $\lambda = \lambda(X_3,t)$. The swelling ratio of the gel can then be expressed as $J = \lambda_0^3\lambda$. As the swelling in the $x_3$-direction is free, the normalized Cauchy stress component $\bar{\sigma}_{33}$ must be zero throughout the gel. From Equation (\ref{eq:cauchy_stress}) we then get
\begin{linenomath}
\begin{equation}
\bar{\sigma}_{33}= Nv\left(\frac{\lambda}{\lambda_0}+\frac{1}{Nv}\left(\ln\frac{J-1}{J}+\frac{1-Nv}{J}+\frac{\chi}{J^2}-\bar{\mu}\right)\right) = 0
\label{eq:cauchy_stress33}
\end{equation}
\end{linenomath}
Equation \ref{eq:cauchy_stress33} yields an expression for the normalized chemical potential that varies through the thickness of the gel as 
\begin{linenomath}
\begin{equation}
\bar{\mu}\left(X_3,t\right)= \frac{Nv\lambda}{\lambda_0}+\ln\frac{J-1}{J}+\frac{1-Nv}{J}+\frac{\chi}{J^2} 
\label{eq:chem_pot}
\end{equation}
\end{linenomath}
From the condition of molecular incompressibility, we can express the nominal concentration of solvent molecules as
\begin{linenomath}
\begin{equation}
C\left(X_3,t\right)= \frac{1}{v}\left(J-1\right)= \frac{1}{v}\left(\lambda_0^3\lambda-1\right)
\label{eq:C}
\end{equation}
\end{linenomath}
By combining Equations (\ref{eq:JK}) and (\ref{eq:MKL}) the nominal flux in the $x_3$-direction can be written as
\begin{linenomath}
\begin{equation}
\begin{split}
\Phi_3= -M_{33}\frac{\partial \mu}{\partial X_3} = -\frac{D}{vkT}\frac{1}{\lambda^2\lambda_0^2}\left(\lambda_0^3\lambda-1\right)\frac{\partial \mu}{\partial X_3}  \\ =  - \frac{D}{v}\left(\frac{\lambda_0}{\lambda}-\frac{1}{\lambda^2\lambda_0^2}\right)\frac{\partial \bar{\mu}}{\partial X_3} 
\end{split}
\label{eq:flux}
\end{equation}
\end{linenomath}
where we can express $\frac{\partial \bar{\mu}}{\partial X_3} $ through use of the chain rule  
\begin{linenomath}
\begin{equation}
\begin{split}
\frac{\partial \bar{\mu}}{\partial X_3} = \frac{\partial }{\partial X_3}\left(\frac{Nv\lambda}{\lambda_0}+\ln\frac{J-1}{J}+\frac{1-Nv}{J}+\frac{\chi}{J^2}\right)   =  \\ \left(\frac{Nv}{\lambda_0} + \frac{1}{\left(\lambda_0^3\lambda-1\right)\lambda}+\frac{Nv-1}{\lambda_0^3\lambda^2}-\frac{2\chi}{\lambda_0^6\lambda^3} \right)\frac{\partial \lambda}{\partial X_3}
\end{split}
\label{eq:dmudx3}
\end{equation}
\end{linenomath}
The conservation of solvent molecules requires that
\begin{linenomath}
\begin{equation}
\frac{\partial C}{\partial t} = -\frac{\partial \Phi\left(X_3,t\right)}{\partial X_3}
\label{eq:cons_c}
\end{equation}
\end{linenomath}
where $\frac{\partial C}{\partial t}$ can be obtained by the use of Equation (\ref{eq:C})
\begin{linenomath}
\begin{equation}
\frac{\partial C}{\partial t} = \frac{\lambda_0^3}{v}\frac{\partial \lambda}{\partial t} 
\label{eq:dcdt}
\end{equation}
\end{linenomath}
Finally, by inserting Equation (\ref{eq:dmudx3}) into Equation (\ref{eq:flux}), we can with the use of Equation (\ref{eq:dcdt}) write Equation (\ref{eq:cons_c}) as
\begin{linenomath}
\begin{equation}
\frac{\partial \lambda}{\partial t} = \frac{D}{\lambda_0^3}\frac{\partial}{\partial X_3}\left(\left(\frac{\lambda_0}{\lambda}-\frac{1}{\lambda^2\lambda_0^2}\right)\left(\frac{Nv}{\lambda_0} + \frac{1}{\left(\lambda_0^3\lambda-1\right)\lambda} +\frac{Nv-1}{\lambda_0^3\lambda^2}-\frac{2\chi}{\lambda_0^6\lambda^3}\right)\frac{\partial \lambda}{\partial X_3}\right)
\label{eq:final_diff}
\end{equation}
\end{linenomath}

\begin{figure}[h]
\centering
  \includegraphics[width=9cm]{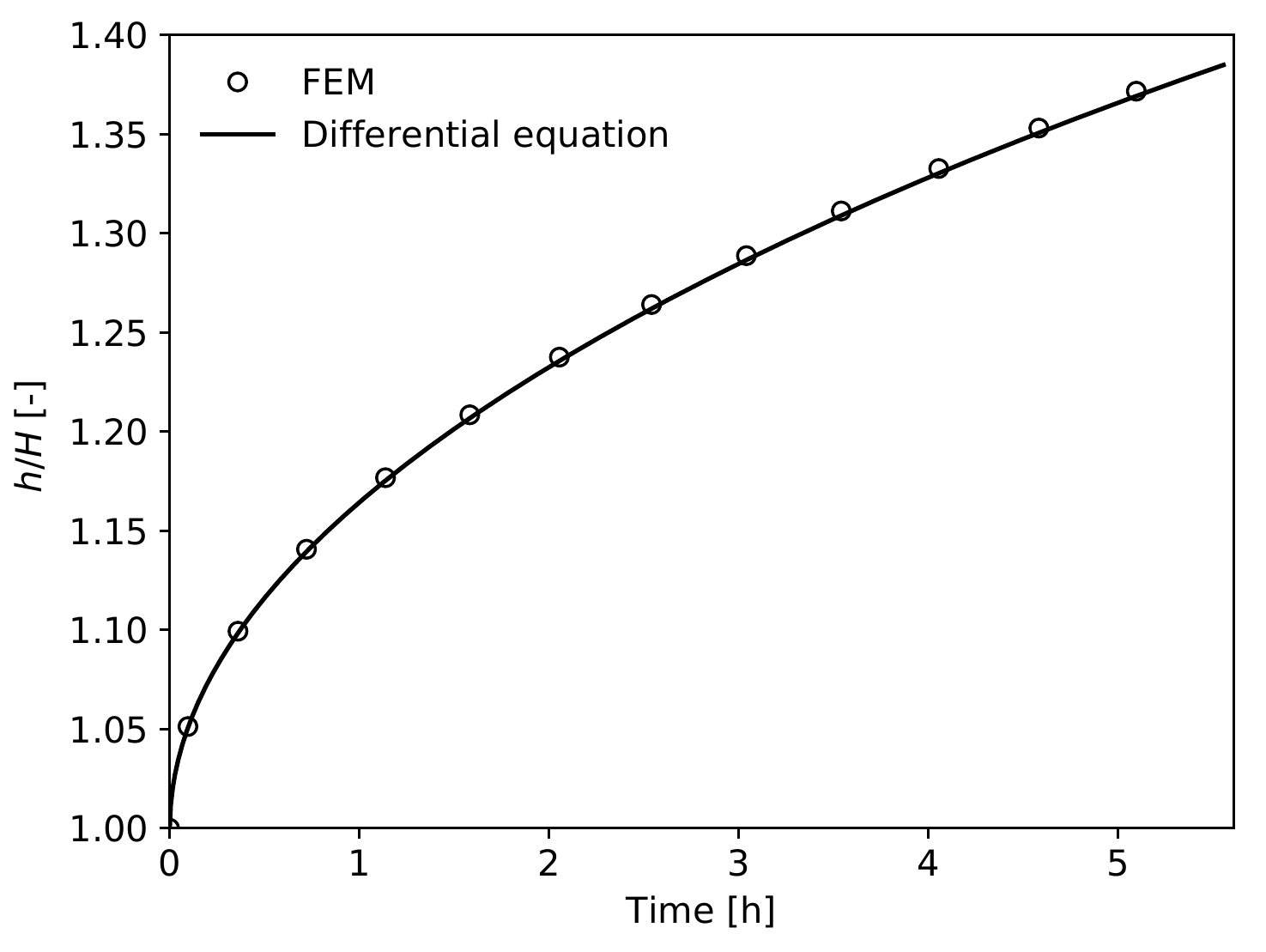}
  \caption{Comparison between the implemented user-subroutine and the solution to Equation (\ref{eq:final_diff}) for transient swelling in a homogeneous plate with Nv = 0.01, $\chi=0.5$, and D=10$^{-11}$ m$^2$/s.}
  \label{f:validate}
\end{figure}

The total out-of-plane swelling ratio $\left(h/H\right)$ obtained through time solving Equation (\ref{eq:final_diff}) with the method of lines is compared to the FEM solution using the implemented Fortran code in Figure \ref{f:validate}, yielding a good validation of the Fortran implementation of the constitutive model.

\subsection{Boundary conditions}
To induce swelling in the hydrogel plate described in Section \ref{sec:prob_def}, the chemical potential is changed at the upper surface from $\bar{\mu}_{1}$ to $\bar{\mu}=0$. To avoid an abrupt change in boundary condition in the finite element simulation, the chemical potential is changed during a smooth step with a duration of 0.2 seconds (assumed to be short compared to the time to buckling initiation for transient swelling). The potential at the exposed surface is then kept constant at zero and the hydrogel swelling is driven by the diffusion process. Through this process, the plate is constrained from in-plane deformation at all lateral faces, while the lower surface is constrained from out-of-plane deformation. To comply with the LPA results and also accommodate for large swelling ratios, we want to mimic a nearly dry state at the start of the simulation, and hence set the initial homogeneous chemical potential to $\bar{\mu}_{1}=-2$, giving an initial free swelling of $J_0 =\lambda_0^3\approx 1.03$.

\subsection{Material parameters}
The material parameters used in this study are given in Table \ref{tab:matpar}. $Nv_s$ defines the normalized stiffness of the substrate, while the parameter $n$ defines the ratio between the stiffness in the film $Nv_f$ and the stiffness of the substrate $Nv_s$ according to $n = Nv_f/Nv_s$. 
For the diffusion coefficient $D$, a value in the range between 10$^{-9}$ and 10$^{-11}$ $\text{m}^2/\text{s}$ is reported as representative for the diffusion of water molecules in a gel \citep{Hong2008,Caccavo2018}. However, we perform simulations for every decade of $D$ in the range from $10^{-11}$ to $1$ $\text{m}^2/\text{s}$ to study the change in the buckling behavior of hydrogels as we gradually go from a transient to an equilibrium swelling process. The set of parameters as given in Table \ref{tab:matpar} leads to a total of 120 simulations to cover the parameter space. 

In addition, simulations were performed for $n=0.1$ and $n=0.8$. These simulations confirm the general trends for soft-on-hard systems as presented in the following. Hence, the results for these two stiffness ratios are not included in the current presentation to enhance readability.  
 \begin{table}[h]
\small
  \caption{\ Material parameters used in the following simulations. $Nv_s$, $n$, and $D$ are changed in each simulation according to the given values or ranges.}
  \label{tab:matpar}
  \begin{tabular*}{0.48\textwidth}{@{\extracolsep{\fill}}llll}
    \hline
    $Nv_s$  & $n$ & $\chi$ & $D \, \left[\text{m}^2/\text{s}\right] $ \\
    \hline
   0.01, 0.001 & 0.5, 2, 5, 20, 100 & 0.5 & $10^{-11}$ - 1   \\
    \hline
  \end{tabular*}
\end{table}

\subsection{Buckling initiation}
\begin{figure}[h]
 \centering
 \includegraphics[width=9cm]{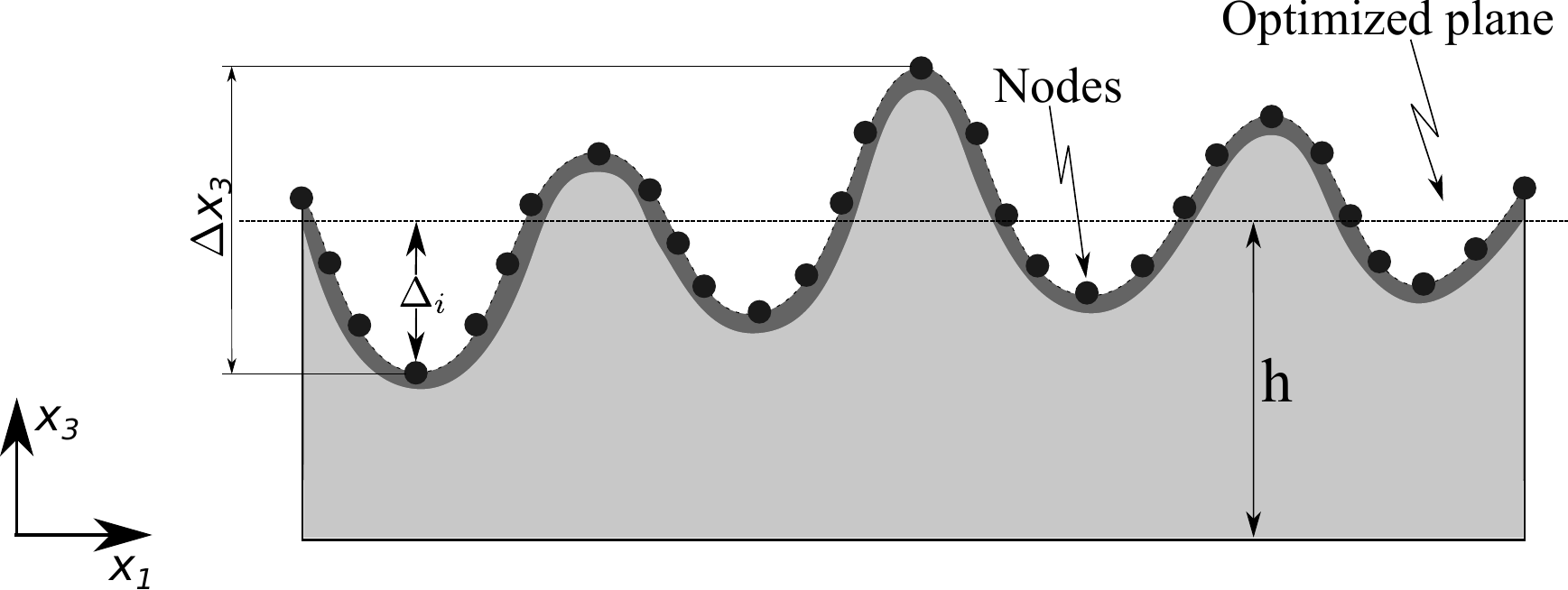}
 \caption{Schematic illustration of plane optimization for a buckled 2D plate}
 \label{f:plane_opt}
\end{figure}

To obtain a measure for the global swelling ratio at the point of buckling and beyond, a plane is optimized with respect to the position of the uppermost layer of nodes in the film as illustrated in Figure \ref{f:plane_opt}. The height of the plane, $h$, is found by minimizing the residual $r$ given as the sum of the squared distances between the nodes in their current position and the plane, i.e. $r=\sum_{i=1}^{n}\left(\Delta_i\right)^2$. The minimization procedure is performed using the SciPy package of Python. The global swelling ratio is defined as the plane height $h$ divided by the initial plate thickness $H= 0.5$ mm.

There are multiple approaches that could be used to quantify the point where buckling initiates in a finite element simulation, e.g. abrupt changes in stress, strain energy or geometry. For practical applications, we consider the topology of the surface to be the most relevant measure. Hence, we define the point of onset of buckling by the use of a threshold value for the difference between the highest and lowest point on the surface, denoted $\Delta x_3$, i.e. buckling occurs if $\Delta x_3  \geq\Delta x_3^{crit}$. Obviously, the obtained results for the onset of buckling will be influenced by the choice of the buckling limit, and the value to be used for $\Delta x_3^{crit}$ should depend on the application of the simulated gel. For the purpose of this study, we choose to use $\Delta x_3^{crit}$=\SI{1}{\micro \meter}. Although the absolute values for the swelling ratio and time to initiation of buckling would be altered with a different threshold value, all trends presented in Section \ref{sec:res} would be preserved if the buckling limit would be set to a lower value like \SI{0.5}{\micro \meter}.

\subsection{Initial imperfection}
To accommodate buckling, an initial imperfection is introduced in the upper surface of the stiff film. A Python script is used to change the $x_3$ position of all the surface nodes in the Abaqus input file. To ensure that the method of introducing this imperfection is not dominating the obtained results, we use two different modes of the initial imperfection. First, we introduce a single sinusoidal wave over the width of the gel
\begin{linenomath}
\begin{equation}
x_3^i=H\left(1 + \epsilon \sin \left(2\pi \frac{x_1^i}{L}\right)\right)
\label{eq:imper_sin}
\end{equation}
\end{linenomath}
where $x_3^i$ is the new $x_3$-position of the node $i$, $H$ is the $x_3$ node coordinate in the perfect model, $\epsilon$ is the maximum perturbation given as a fraction of the initial height, and $x_1^i$ is the $x_1$-position of node $i$. Second, we use a random perturbation given by 
\begin{linenomath}
\begin{equation}
x_3^i=H\left(1+ \epsilon p\right)
\label{eq:imper_rand}
\end{equation}
\end{linenomath}
where $p \in\left[-1,1\right]$ is a pseudorandom number. While the expression in Equation (\ref{eq:imper_sin}) gives a smooth surface, it assumes a specific shape of the surface of the gel. The random imperfection in Equation (\ref{eq:imper_rand}), on the other hand, avoids \textit{a priori} assumptions of the shape of the initial surface but can produce highly uneven surfaces.

\begin{figure}[h]
 \centering
 \includegraphics[width=9cm]{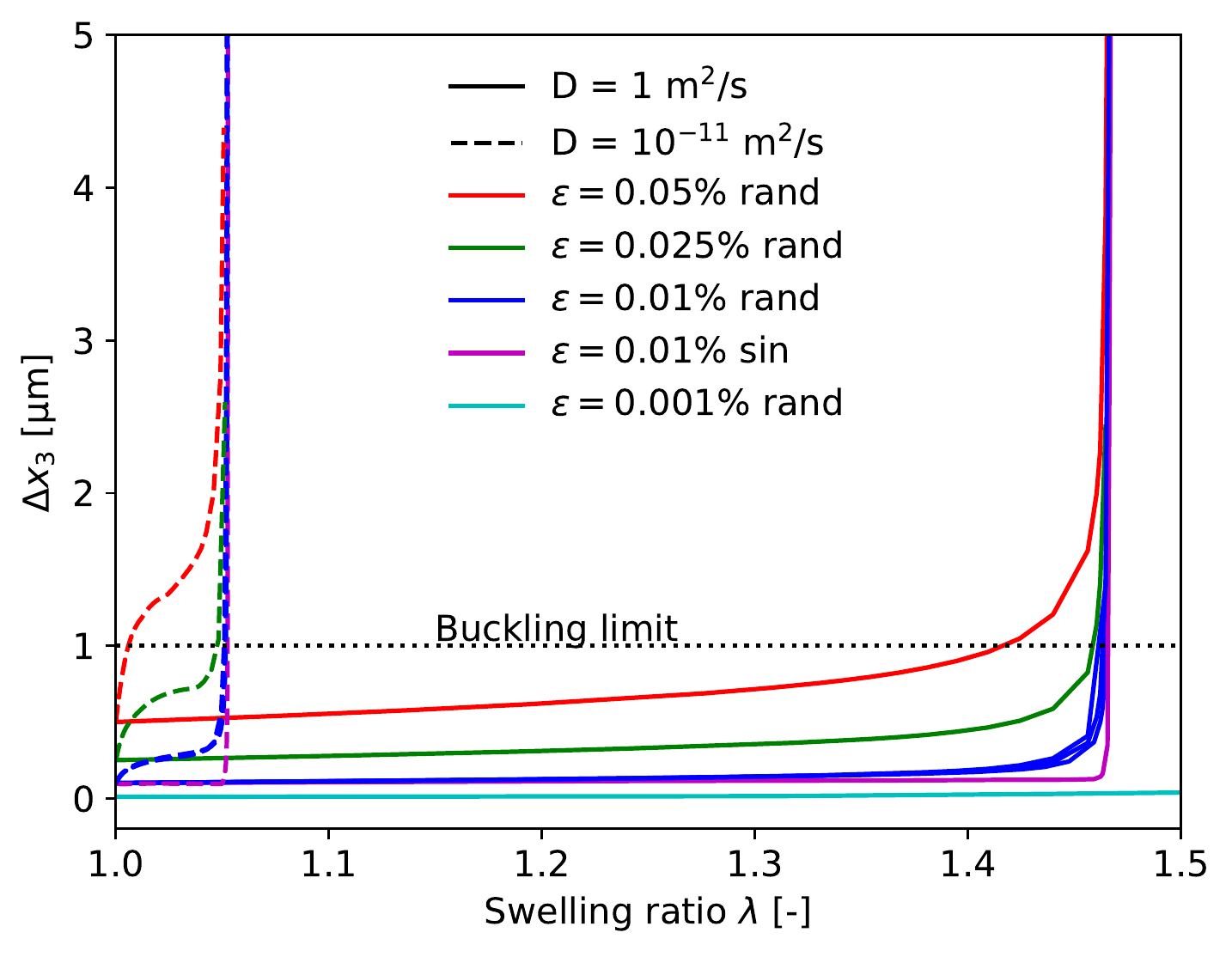}
 \caption{Effect of the size of the initial imperfection for the evolution of $\Delta x_3$ vs swelling ratio, using $Nv_s=0.01$ and $n=5$.}
 \label{f:effect_imper_size}
\end{figure}

The effect of the initial imperfection size, $\epsilon$, on the maximum height difference in the plate surface, $\Delta x_3$, for slow and fast diffusion processes, $D=10^{-11}$ and $D=1$ m$^2$/s, respectively, is shown in Figure \ref{f:effect_imper_size}. Clearly, a larger imperfection size would lead to faster growth of the height difference in the plane, however, the same asymptote is reached for all simulations where buckling is initiated. It can be seen that for the random initial imperfections, there is a significant increase in $\Delta x_3$ at the beginning of the simulations with a slow diffusion process. This effect arises as the random imperfection introduces peak and valley nodes in the mesh, where the peak nodes would be exposed to more solvent and hence swell faster than the valley points (similar to free swelling of a cube where the corners would swell faster than the center of the faces \cite{Zhang2009}). This initial effect is not seen for $D=1$ m$^2$/s, as this would resemble an equilibrium process with equal swelling ratios in both peak and valley points, nor for the sinusoidal imperfection as this surface will be initially smooth. To demonstrate the effect of the randomness introduced in the mesh by Equation (\ref{eq:imper_rand}), results from three subsequent random imperfections applied to an initially perfect mesh using $\epsilon=0.01\%$ are shown (i.e. overlapping blue curves in Figure \ref{f:effect_imper_size}). Finally, the plot also shows that the imperfection must be larger than a critical size to trigger buckling, as seen for the curve for a random imperfection with $\epsilon=0.001\%$ where no buckling is initiated. 

In all further calculations, we set $\epsilon$ to be 0.01\%, meaning that the plate height at each node will be in the range between 0.49995 mm and 0.50005 mm and that we initially have $\Delta x_3 \le$ \SI{0.1}{\micro \meter}. It is important to note that the initial imperfection must be smaller than the threshold value used to quantify the onset of buckling. In the following, we perform simulations using either the harmonic imperfection (Equation (\ref{eq:imper_sin})) or the random imperfection (Equation (\ref{eq:imper_rand})). 

\subsection{Plane strain assumption}
To capture both the gradient of the chemical potential through the thickness of the plate in the transient swelling analyses and the evolving in-plane buckling pattern requires a relatively fine mesh, putting significant demands on computational resources.

However, it can be noted that for the problem at hand, a plane strain assumption would be correct up to the point of buckling. After buckling has initiated though, a plane strain assumption would restrain the model to a 1D buckling pattern, in contradiction with experimental and theoretical studies \cite{Cai2011a} having shown that other buckling patterns, like a hexagonal or herringbone mode, would be energetically favorable. 

\begin{figure}[h]
\centering
  \includegraphics[width=14cm]{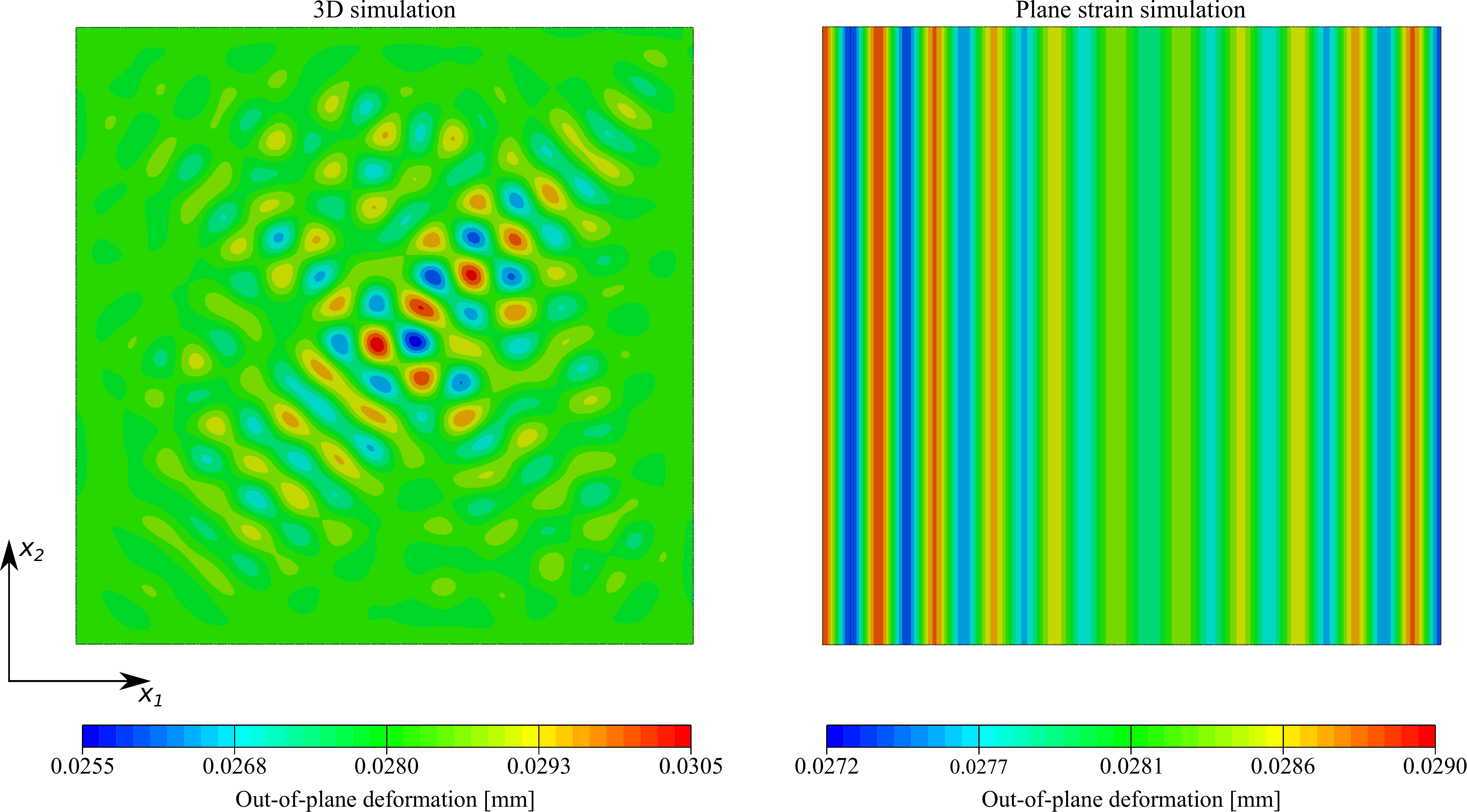}
  \caption{Illustration of buckling mode for an equilibrium simulation with $Nv_s=0.001$ and $n=100$ obtained in a 3D and a 2D simulation. The fringe plot gives the out-of-plane deformation. The 2D result is extruded in the $x_2$-direction to visualize its 3D resemblance.}
  \label{f:2d_3d_mode}
\end{figure}

\begin{figure}[h]
\centering
  \includegraphics[width=9cm]{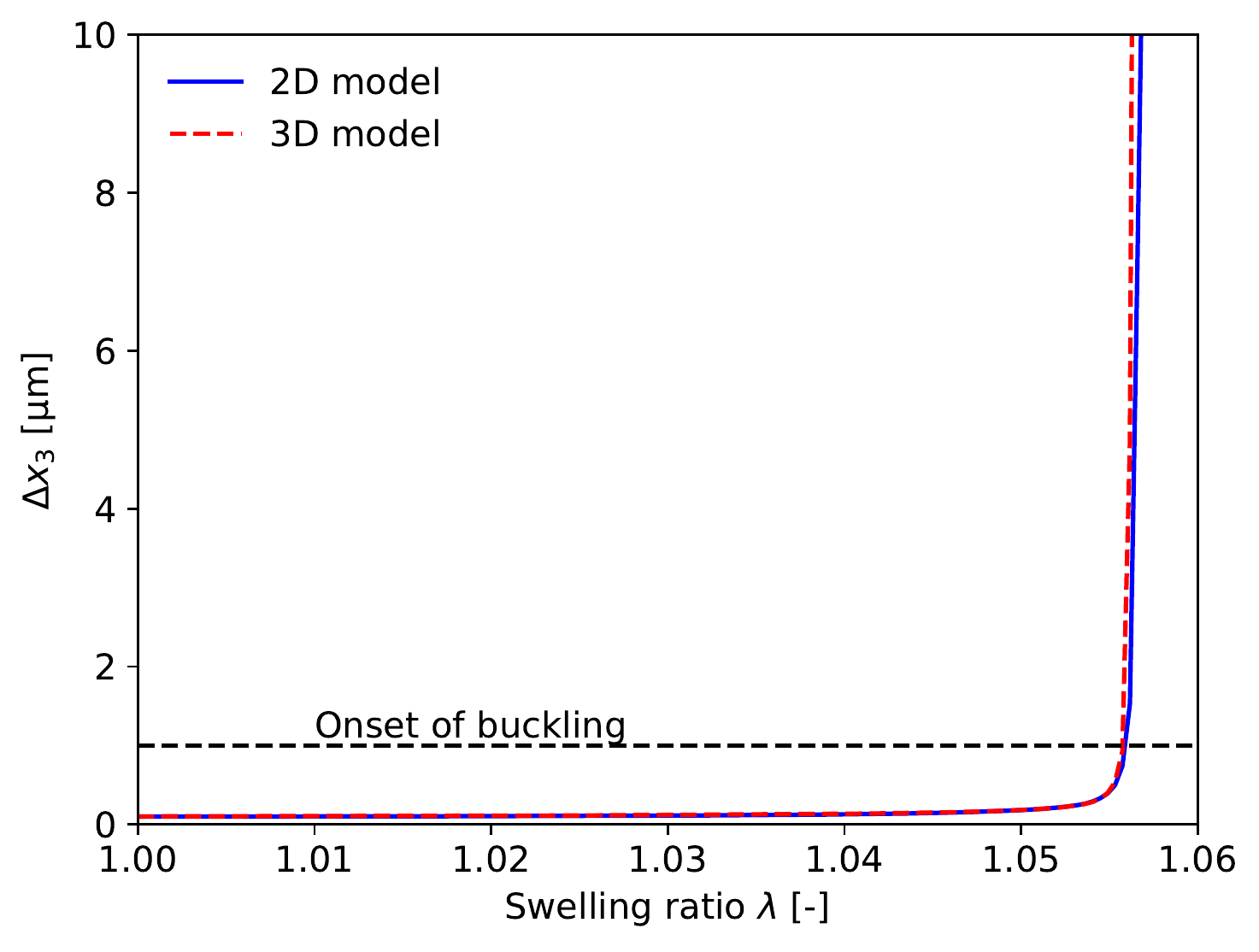}
  \caption{Comparison of the predicted onset of buckling for a 2D and a 3D model during equilibrium swelling.}
  \label{f:2d_3d_dz}
\end{figure}

To investigate if a 2D plane strain assumption would predict buckling at the same swelling ratio as a full 3D model, despite the difference in bucking mode, a simple case of equilibrium swelling with $Nv_s=0.001$ and $n=100$ was used. Due to the homogeneous chemical potential and the large wavelength of the resulting buckling pattern, a relatively coarse mesh could be used. The 2D and 3D problems were both discretized in the same manner through the thickness using fully integrated higher order elements. The obtained buckling patterns and the increase in $\Delta x_3$ during swelling are shown in Figures \ref{f:2d_3d_mode} and \ref{f:2d_3d_dz} respectively. The modes shown in Figure \ref{f:2d_3d_mode} are obtained from the first increment after initiation of buckling according to the threshold level $\Delta x_3  \geq$ \SI{1}{\micro \meter}. The initial buckling pattern in the 3D simulation can be seen to have a checkerboard pattern in the central region of the plate, this gradually expands to cover the plate before it would transition to a herringbone mode at larger swelling ratios \cite{Cai2011a}. For the plane strain simulations, the 2D nature of the model forces the plate to buckle in the less advantageous 1D buckling pattern more commonly observed when one of the in-plane stresses dominates \cite{Breid2011}. From the fringe plots, it can be seen that the value of $\Delta x_3$ is larger in the 3D simulation. However, from Figure \ref{f:2d_3d_dz} it is seen that the swelling ratio (i.e. the height of the optimized plane) at the onset of buckling is similar between the 2D and 3D simulations. Hence, to reduce the computational demands, a plane strain model is used in the further to study the point of onset of instability. This limits the study from considering detailed post-buckling analysis as the correct buckling pattern cannot be obtained. 

\subsection{Element type and size}
For the further 2D simulations, the plate is discretized by fully integrated linear plane strain temperature-displacement elements (denoted CPE4T in Abaqus) where the temperature field is used to mimic the distribution of the normalized chemical potential through the plate (see Appendix \ref{app:heat_ana} for the analogy between gel diffusion and heat transfer). Linear elements were preferred over second-order elements as the first use a lumped heat capacity matrix and hence avoid spurious oscillations in the temperature field during changes in the boundary conditions \cite{bergheau2013finite}. Thereby, with our use of the heat transfer and gel swelling analogy, oscillations in the chemical potential during transient swelling is avoided. See  \cite{Bouklas2015,Dortdivanlioglu2018} and the references therein for a further discussion on oscillations in the chemical potential during finite element simulations of transient gel swelling.

\begin{figure}[h!]
\centering
  \includegraphics[width=15cm]{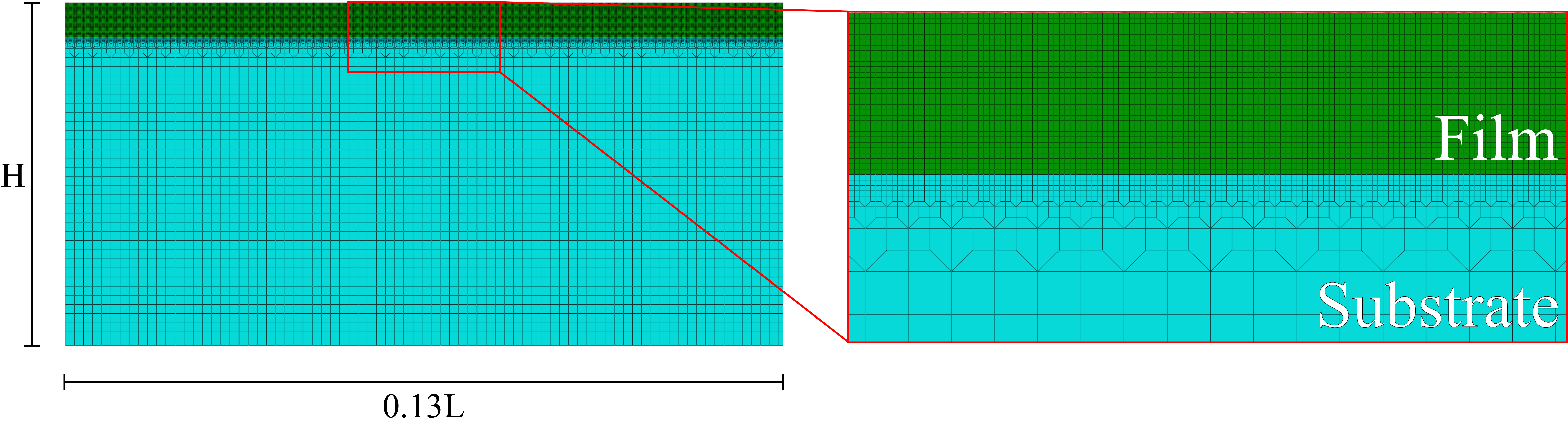}
  \caption{Final mesh used in simulations. Note that while the full height H is shown in the left figure, only 13\% of the length L is included for improved readability.}
  \label{f:plate_mesh}
\end{figure}

\begin{figure}[h!]
\centering
  \includegraphics[width=9cm]{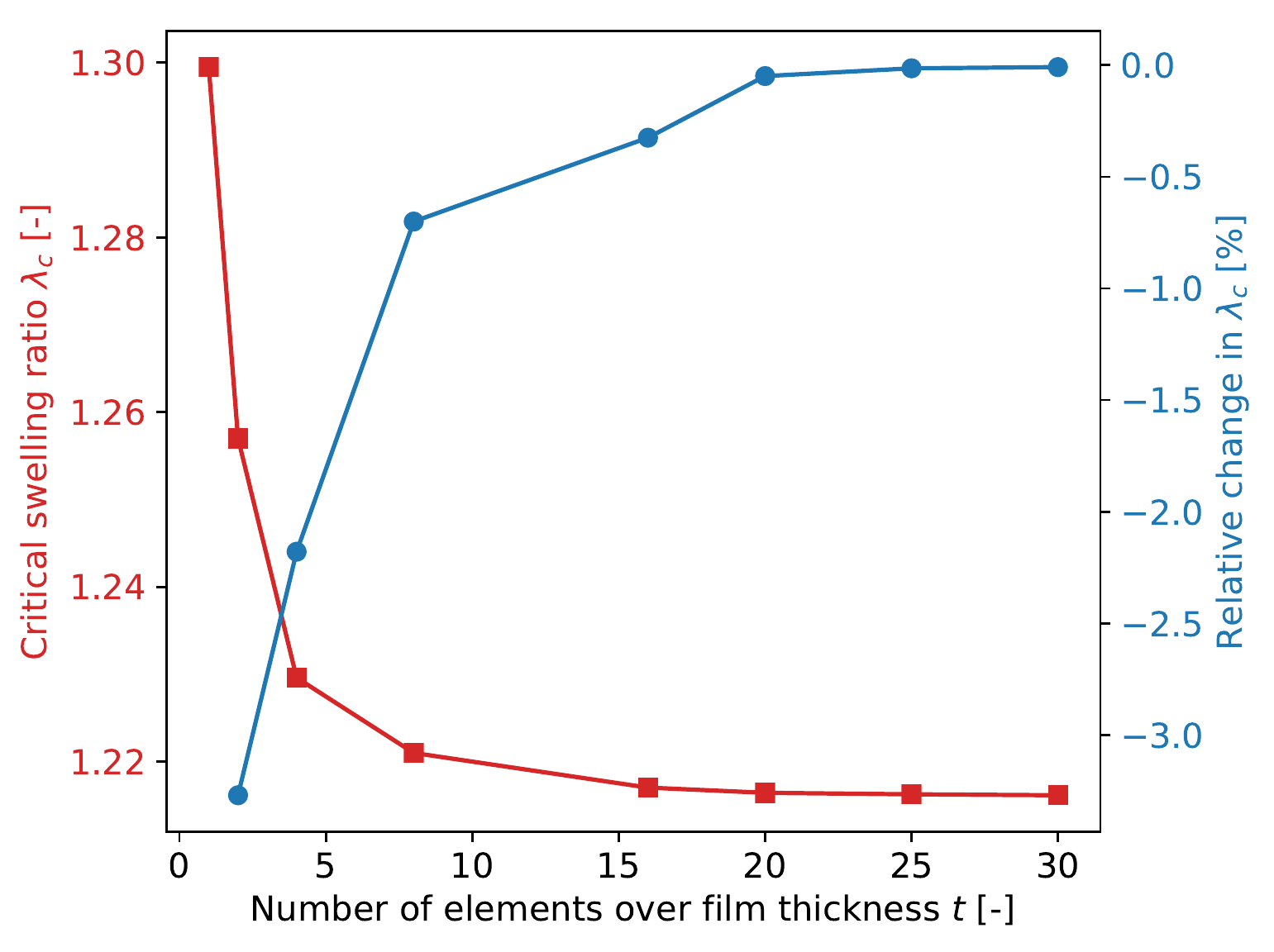}
  \caption{Effect of mesh refinement for transient swelling with $Nv_s=0.01$,  $n=2$, and $D=10^{-11}$ m$^2$/s. The red squares refer to the swelling ratio, while the blue circles refer to the relative change in the swelling ratio compared with the previous (larger) element size.}
  \label{f:element_size}
\end{figure}

The effect of the element size on the obtained results was studied by running simulations with $Nv_s=0.01$, $n=2$ and $D=10^{-11}$ m$^2$/s using between 1 and 30 square elements over the film thickness. To avoid a change in the surface topology as the mesh is refined, the harmonic imperfection (Equation (\ref{eq:imper_sin})) was used. As the gradient in the chemical potential is largest close to the surface of the gel, the total size of the model was reduced by gradually increasing the mesh size going from the film and into the substrate as illustrated in the enlarged view in Figure \ref{f:plate_mesh}. The resulting swelling ratio at the onset of instability as a function of the number of elements over the thickness of the stiff film is shown in Figure \ref{f:element_size}. The red squares refer to the swelling ratio, while the blue circles refer to the relative change in the swelling ratio compared with the previous (larger) element size. The same trend is observed with respect to the time to onset of instability. As the swelling ratio at the onset of buckling seems to be converged for a mesh with 30 square elements over the film thickness, this mesh size, leading to a total of 194 400 2D elements, is used in all further calculations. It is worth noting that for this element size, the random imperfection with $\epsilon$ set to 0.01\% produces a maximum perturbation of the upper nodes of 3\% of the initial element height. A sketch of the final mesh with a random imperfection can be seen in Figure \ref{f:plate_mesh}. 

\section{Results and discussion} \label{sec:res}
\subsection{LPA vs FEM for equilibrium swelling} \label{sec:res_LPA}
\begin{figure}[h]
	\centering
	\includegraphics[width=9cm]{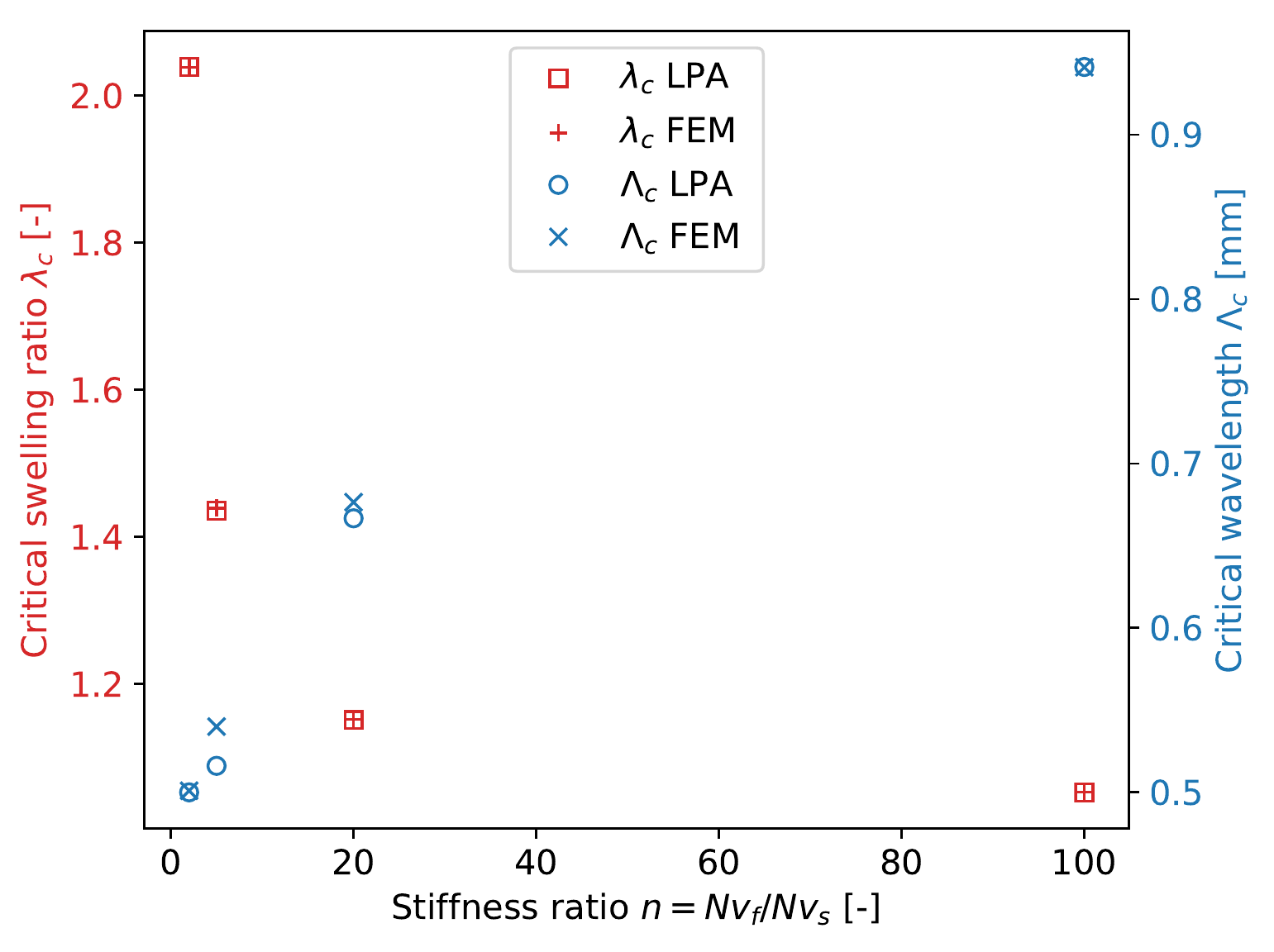}
	\caption{Comparison between LPA and FEM results for hard-on-soft systems. The FEM results are obtained using $Nv_s=0.001$ and $D=1$ m$^2$/s. The LPA results are scaled such that the number of waves over the length of the plate $L$ yields a whole number. Red squares and plusses refer to the left ordinate, while blue circles and crosses refer to the right ordinate.}
	\label{f:com_fem_lpa}
\end{figure}

 A comparison between the FEM and the LPA results for hard-on-soft systems during equilibrium swelling are shown in Figure \ref{f:com_fem_lpa} where the left ordinate indicates the critical swelling ratio at buckling initiation and the right ordinate indicates the wavelength of the buckling pattern at initiation. The FEM results are shown for $Nv_s = 0.001$ and  $D=1$ m$^2$/s. For the critical swelling ratio, a nearly perfect correspondence can be observed between the two analysis methods. The predicted wavelength, on the other hand, shows a perfect correspondence between the two methods for $n=2$ and $n=100$, for the two other stiffness ratios a reasonably good correspondence is seen. A reason for the discrepancy in the wavelength results can stem from the finite width of the FEM model, while the LPA method assumes an infinitely wide plate. However, the generally good correspondence between the FEM and LPA results indicates that the length of the plate $L$ in the FEM model is sufficient to yield results representative for infinitely wide plates. It can be noted that a comparison between FEM and LPA results using $Nv_s = 0.01$ shows a similar correspondence, and were omitted from Figure \ref{f:com_fem_lpa} merely to enhance readability. 
 
For the soft-on-hard system ($n=0.5$) both the LPA and the FEM approaches predict creasing as the initial instability mode. For the swelling ratio at the onset of buckling, on the other hand, the FEM and LPA results deviate with about 8\%. However, it is observed that for $n=0.5$ and $Nv_s = 0.001$ the FEM results have not fully reached the equilibrium plateau level at the diffusion coefficient of 1 m$^2$/s (seen in Section \ref{sec:res_ratio}, Figure \ref{f:Effect_plane}), explaining a part of the discrepancy.

\subsection{Stress and chemical potential profile at the initiation of buckling }\label{sec:stress_chem_profile}
\begin{figure}[h!]
	\centering
    	\begin{subfigure}[t]{0.43\textwidth}
		\includegraphics[width=\textwidth]{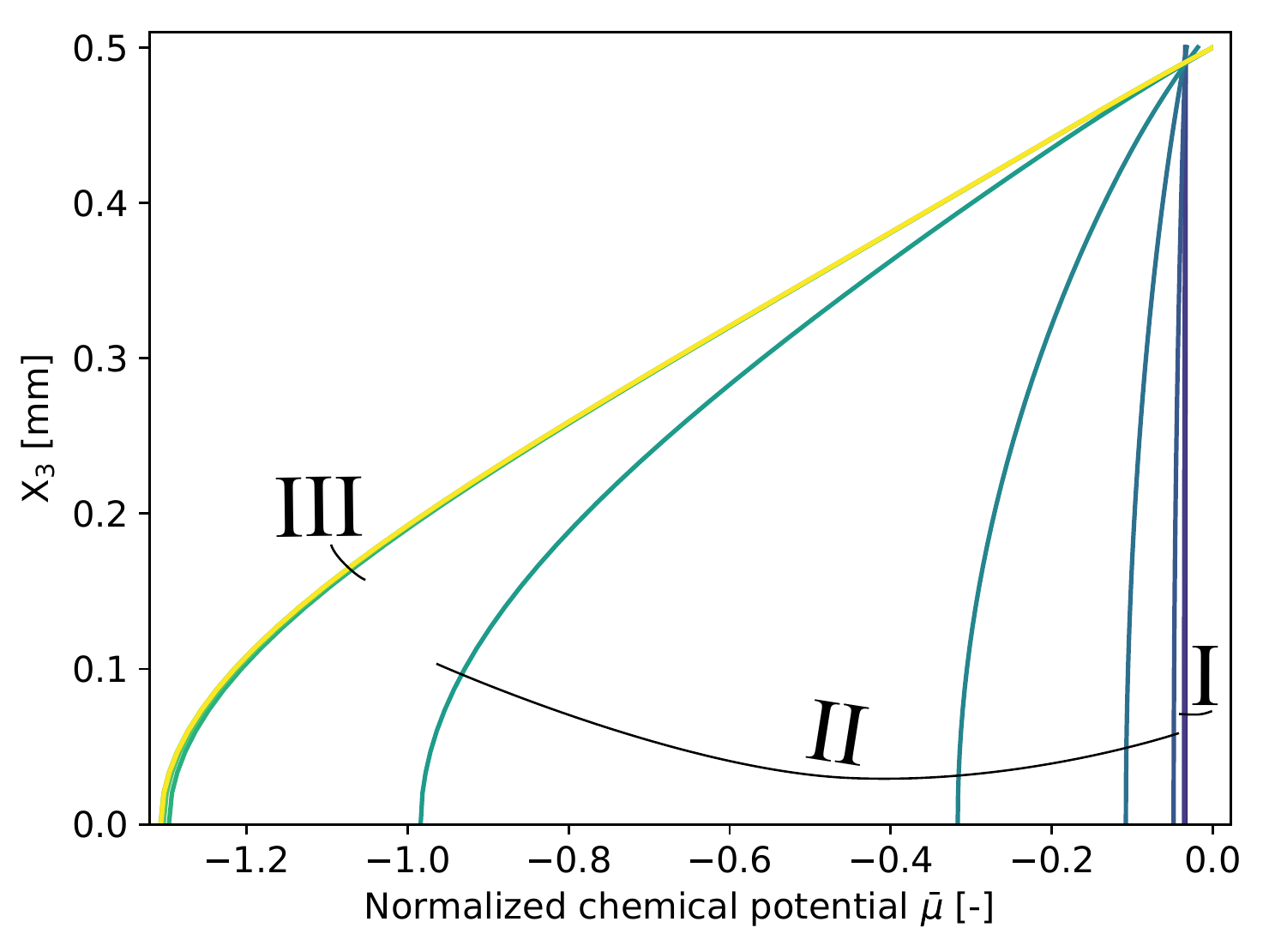}
		\caption{}
         \label{f:chem_prof}
	\end{subfigure}
	\begin{subfigure}[t]{0.43\textwidth}
		\includegraphics[width=\textwidth]{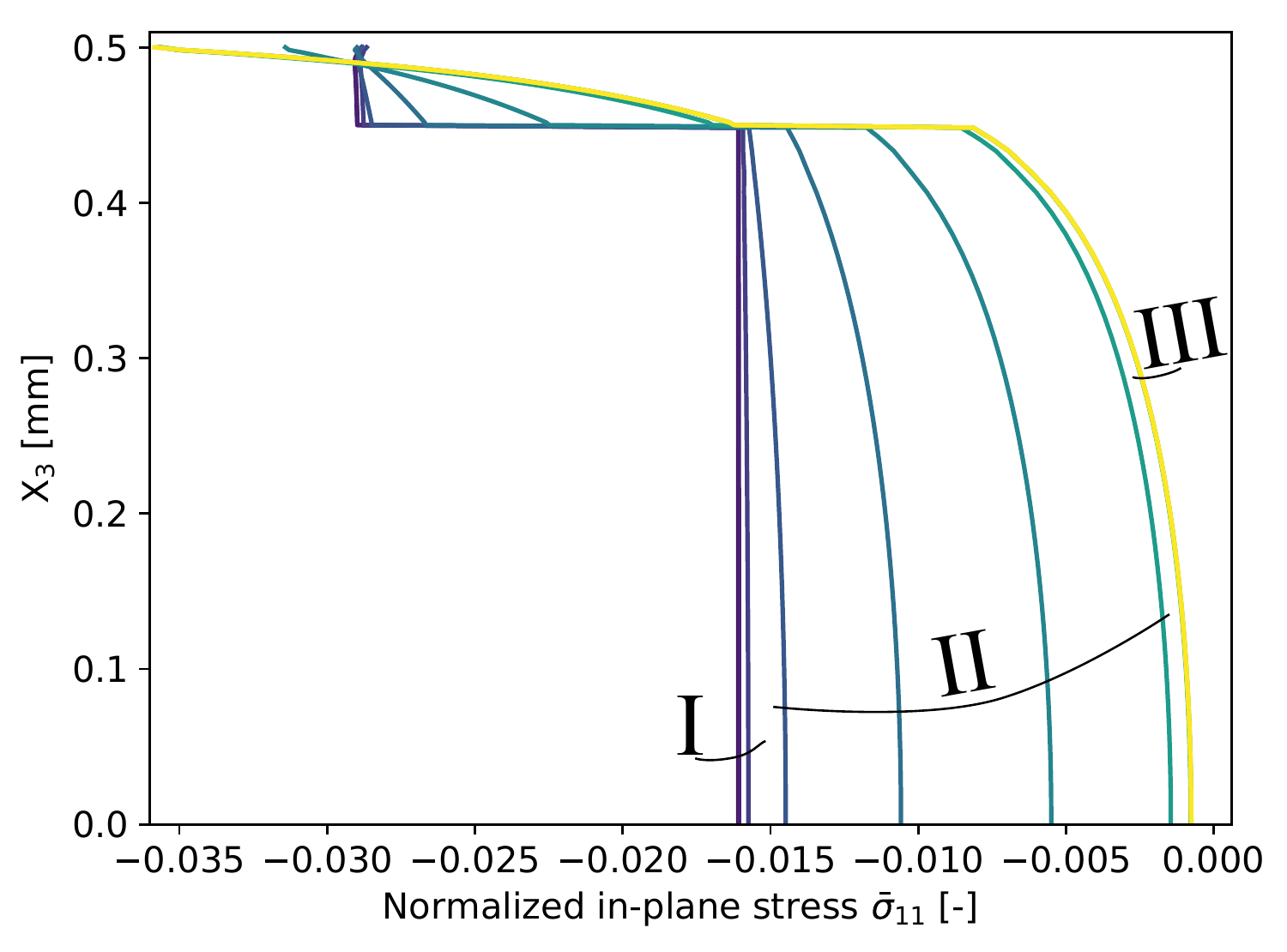}
		\caption{}
        \label{f:stress_prof}
	\end{subfigure}
	\begin{subfigure}[t]{0.09\textwidth}
		\includegraphics[width=\textwidth]{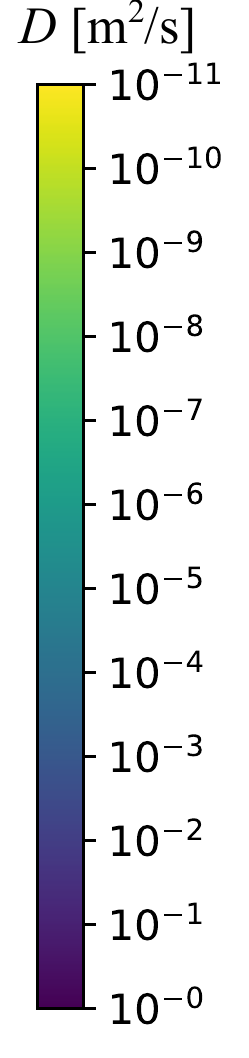}
	\end{subfigure}
  \caption{Distribution of normalized chemical potential (a) and normalized horizontal stress (b) through the initial thickness of the plate at the onset of buckling for all values of the diffusion coefficient $D$. Curves obtained for $Nv_s=0.01$ and $n=2$. Similar trends can be found for other stiffness parameters.}
  \label{f:chem_stress_profile}
\end{figure}

The distribution of the normalized chemical potential, $\bar{\mu}$, and the normalized in-plane stress, $\bar{\sigma}_{11}$, through the thickness of the plate at the onset of buckling are shown in Figure \ref{f:chem_stress_profile} for all values of the diffusion coefficient using stiffness parameters of $Nv_s=0.01$ and $n=2$ (note that the presented trends are representative also for other parameter combinations). Here, $X_3$ is used to denote the $x_3$-position in the plate in the intermediate state (i.e. the starting point for the simulations) such that the curves are easily comparable. The data are extracted from the right edge of the plane strain plate, however, the profiles would be representative for the whole plate at the initiation of buckling. 

For the normalized chemical potential shown in Figure \ref{f:chem_prof}, three distinct regions of the diffusion coefficient can be found. First, for all simulations with $D\ge10^{-2}$ m$^2$/s, we see that the chemical potential at the onset of buckling is nearly homogeneous through the thickness of the plate and we have overlapping curves, i.e. equilibrium swelling is represented. We denote this region I. Second, we have a transition region in the range $10^{-6}$ m$^2$/s $\le D\le10^{-3}$ m$^2$/s where the chemical potential through the plate is increasingly inhomogeneous as the diffusivity is reduced. We denote this region II. Finally, the results for a diffusion coefficient $D\le10^{-7}$ m$^2$/s produces overlapping curves as the profile of the chemical potential through the thickness at the onset of buckling is nearly independent of the diffusion coefficient. We denote this region III. 

For the profiles of the normalized in-plane stresses at the onset of buckling seen in Figure \ref{f:stress_prof}, we see the same three regions as for the normalized chemical profile, with overlapping curves for fast (I) and slow (III) diffusion and a transition zone in-between (II). For the equilibrium swelling simulations (i.e. $D\ge10^{-2}$ m$^2$/s), the chemical potential is homogeneous through the thickness of the plate, and consequentially the compressive stress changes through the plate only due to the stiffness difference between the film and the substrate. 

For the slow diffusion processes (i.e. $D\le10^{-7}$ m$^2$/s), on the other hand, the compressive stress is gradually reduced (i.e. becomes less negative) going downwards through the film thickness. Then there is a discontinuous reduction as one goes from the stiff film to the soft substrate. In the substrate, there is a further gradual decrease towards zero as one approaches the bottom of the plate. It can also be seen that the maximum compressive stress, located at the top of the film, increase as the diffusion coefficient is reduced.

\subsection{Swelling ratio at buckling initiation}\label{sec:res_ratio}
\begin{figure}[h!]
\centering
  \includegraphics[width=9cm]{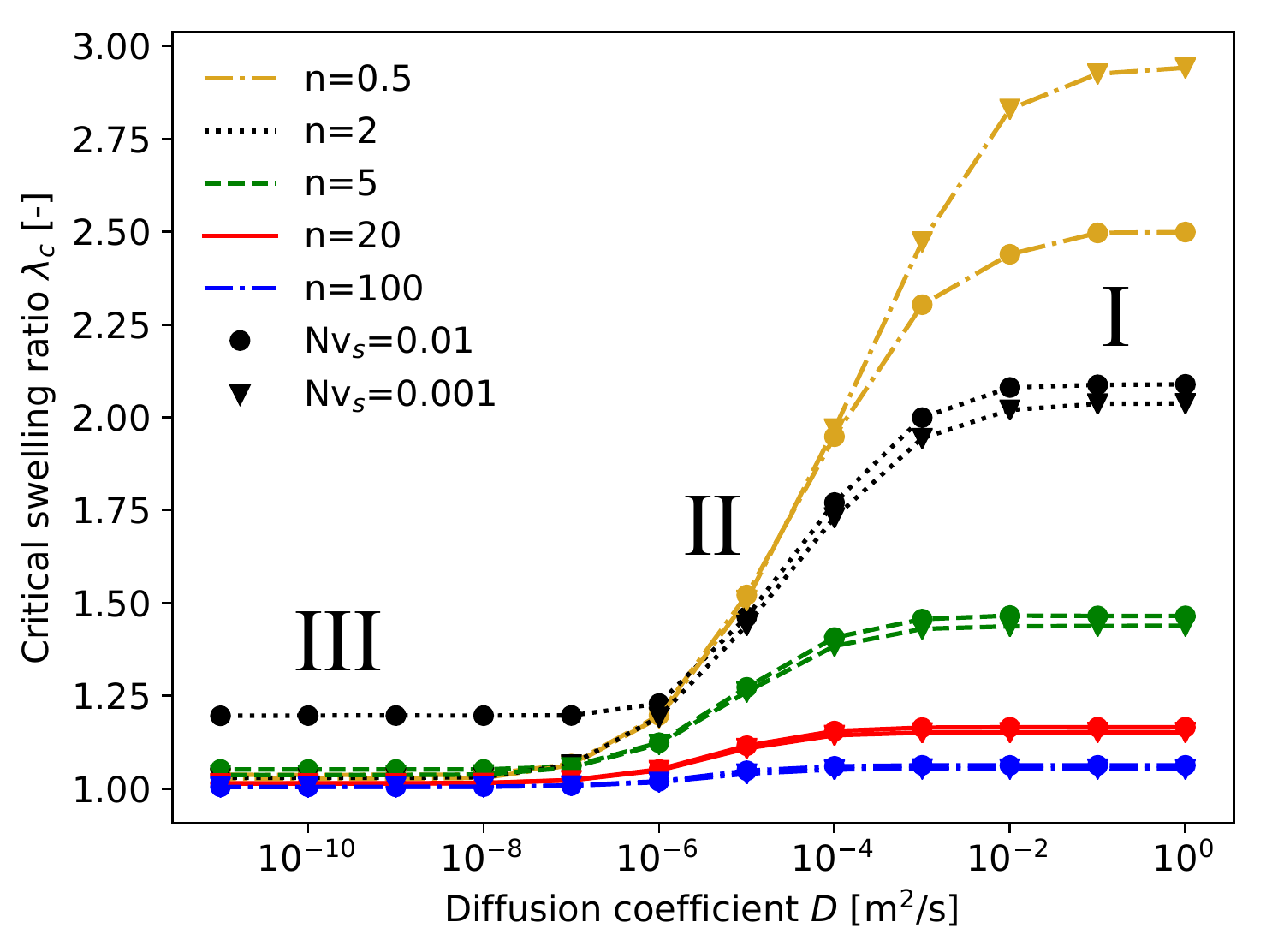}
  \caption{Effect of the diffusion coefficient $D$ on the critical swelling ratio for all parameter combinations.}
  \label{f:Effect_plane}
\end{figure}

Figure \ref{f:Effect_plane} shows the swelling ratio of the plate (i.e. the global amount of swelling $h/H$) at the onset of buckling for values of the diffusion coefficient $D$ in the range from $10^{-11} $ to $1 \,  \text{m}^2/\text{s}$, $n$ between 100 and 0.5, and a normalized stiffness of the substrate of 0.001 or 0.01. The plotted data were obtained using a random initial imperfection in the upper surface of the plate (i.e. Equation \ref{eq:imper_rand}), however, a similar plot would be produced if a harmonic initial imperfection was used (i.e. Equation \ref{eq:imper_sin}). In accordance with the results for the normalized chemical potential and compressive stress profiles at the onset of buckling (discussed in Section \ref{sec:stress_chem_profile}) the swelling ratio at the onset of buckling can be divided into the three distinct regions I, II, and III. In region I, there is a relatively large and stable swelling ratio at buckling initiation that equals the equilibrium solution. Then, in region II there is a transition zone where the swelling ratio at buckling initiation is gradually reduced as the diffusivity is lowered. Finally, in region III a relatively low and stable swelling ratio at buckling initiation is obtained. In the plot, it can also be seen that the range of the diffusion coefficient in which the three regions occur depends on the value of $n$, with the transitions between the regions moving towards higher values of $D$ as $n$ is reduced. 

Further, the swelling ratio at the onset of instability can also be seen to depend on the absolute stiffness of the substrate and the film, and not only the ratio between the two. For the soft-on-hard system ($n=0.5$), a larger critical swelling ratio is obtained at high values of $D$ for the softer gels (i.e. $Nv_s = 0.001$) compared to the stiffer gels (i.e. $Nv_s = 0.01$). For hard-on-soft gels ($n>1$) on the other hand, a larger swelling ratio at the onset of buckling is obtained for the stiffer gels (i.e. $Nv_s = 0.01$) compared to the softer gels (i.e. $Nv_s = 0.001$). This is especially seen in the results for $n=2$ combined with low values of $D$, where a swelling ratio close to unity is present at the onset of buckling if $Nv_s=0.001$, considerably below a swelling ratio of 1.2 as found for $Nv_s=0.01$. A further discussion on the large difference between these parameter combinations is given in Section \ref{sec:buc_pattern} discussing the obtained buckling profiles. 

\subsection{Time to buckling initiation}\label{sec:time_initiation}
\begin{figure}[h]
\centering
  \includegraphics[width=9cm]{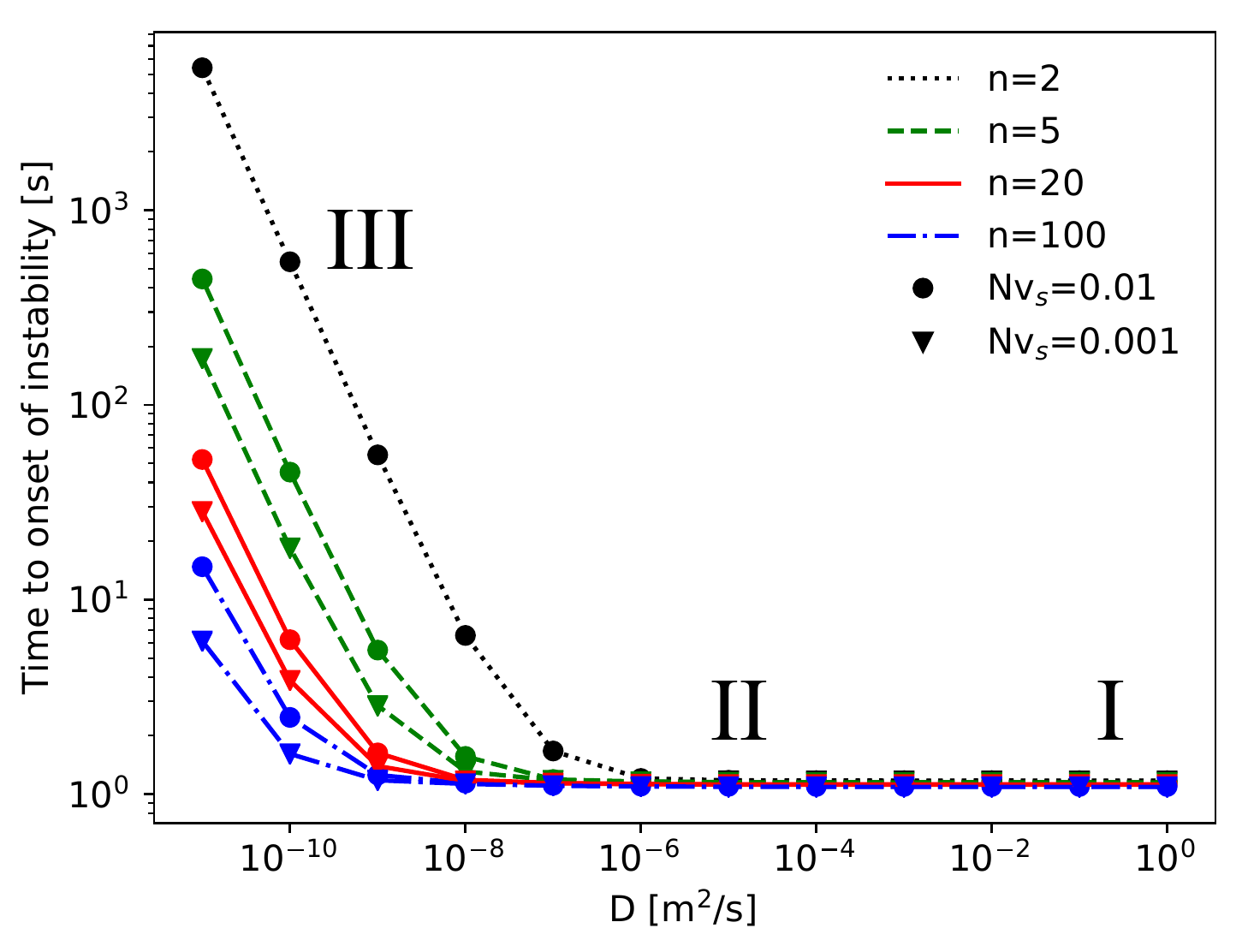}
  \caption{Effect of the diffusion coefficient $D$ for the time to instability for all parameter combinations.}
 \label{f:Effect_time}
\end{figure}

Figure \ref{f:Effect_time} shows the time to the onset of instability on a log-log plot for the simulations where wrinkling was obtained as the mode of instability (see Section \ref{sec:buc_pattern} for a discussion on buckling profiles). Simulations predicting creasing were excluded from the plot as the time to onset of creasing is known to show a strong mesh dependence \cite{Bouklas2015}. 

In Figure \ref{f:Effect_time}, it is seen that there is a negligible effect on the time to the onset of buckling in region I and II (i.e. $D\ge10^{-6}$ m$^2$/s), however, reducing the diffusion coefficient even further (i.e. into the physically reasonable region for water molecules diffusing in a gel) gives a clear rise in the time needed for buckling to occur, with an increasing effect as $n$ is reduced for hard-on-soft gels. For all parameter combinations, a linear log-log relation between the time to instability and the diffusion coefficient can be seen in region III, corresponding to the stable region for swelling at the onset of instability for low values of $D$ as shown in Figure \ref{f:Effect_plane}. It is also seen that there is a clear dependence on the absolute value of the stiffness in the plate and the substrate and not only the ratio between the two, with longer times to buckling initiation for stiffer gels. 

From the time to instability results it can be noted that the time used on the smooth increase of the chemical potential at the surface of the gel in the FEM model, i.e. 0.2 seconds, is negligible compared to the time to the onset of buckling for low values of the diffusivity. In addition, with the timescale needed to obtain buckling in region III, it is assumed that the contribution from viscoelastic effects in the polymer network would be negligible. For region I and II, on the other hand, further studies are needed to investigate how viscous effects might alter the initiation of buckling.   

\subsection{Buckling profiles}\label{sec:buc_pattern}
\begin{figure}[h!]
\centering
  \includegraphics[width=9cm]{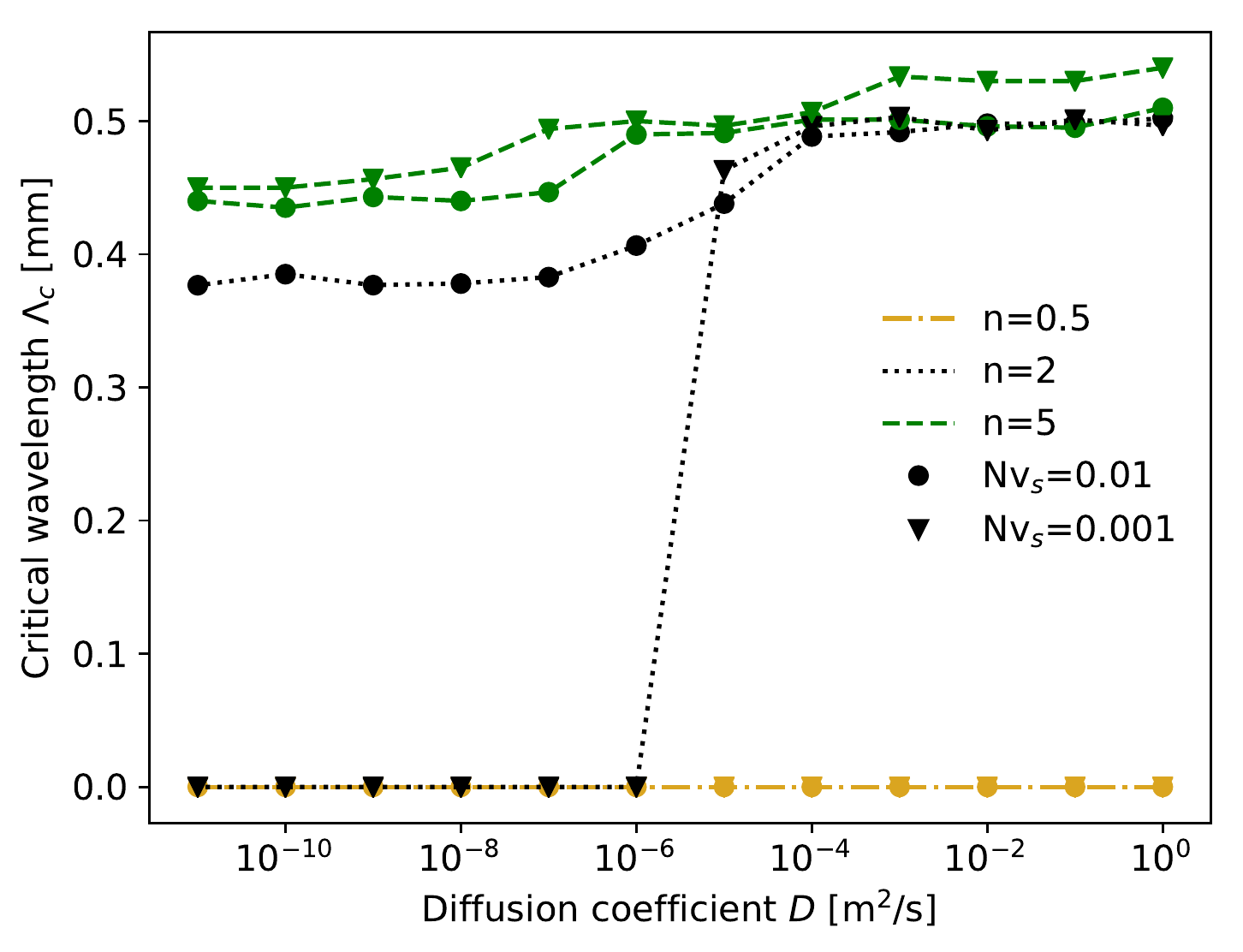}
  \caption{Effect of the diffusion coefficient $D$ for the critical wavelength for $n=0.5$, $n=2$ and $n=5$.}
 \label{f:Effect_D_wl}
\end{figure}

 To study the dependence on the diffusion coefficient for the obtained buckling pattern we extract the wavelength found in the FEM simulations for all values of $D$. Figure \ref{f:Effect_D_wl} shows the obtained wavelengths for $n=0.5$, $n=2$ and $n=5$. For $n=20$ and $n=100$, no clear trend in the critical wavelength could be seen.
 
\begin{figure}[h!]
	\centering
	\begin{subfigure}[t]{0.9\textwidth}
		\includegraphics[width=\textwidth]{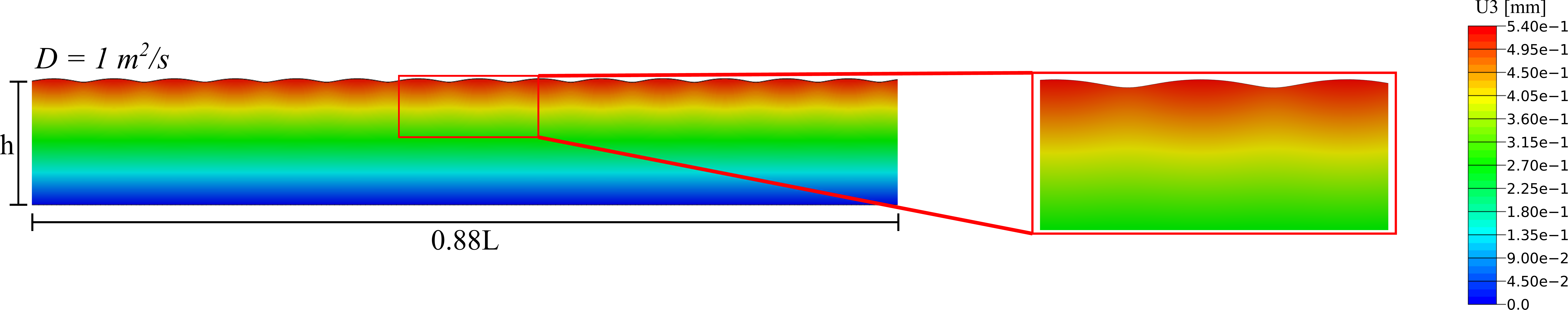}
		\caption{}
	\end{subfigure}
	\begin{subfigure}[t]{0.9\textwidth}
		\includegraphics[width=\textwidth]{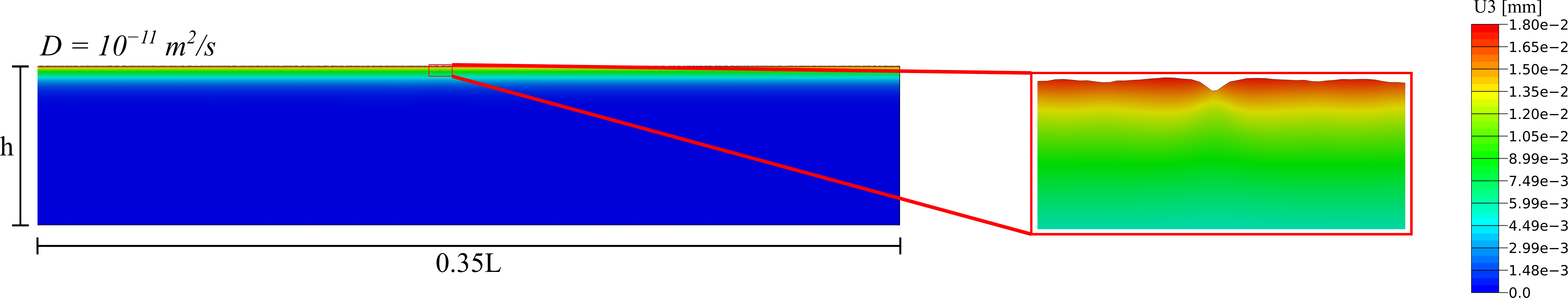}
		\caption{}
	\end{subfigure}
  \caption{Buckling profiles obtained using diffusion coefficient $D=1$ m$^2$/s (a) and $D=10^{-11}$ m$^2$/s (b), for $Nv_s=0.001$ and $n=2$. While the full height in the swollen configuration, $h$, is shown in the left picture, only a part of the width is included. The fringe plot denotes the deformation in the $x_3$ direction $U3$.}
  \label{f:buckling_profile}
\end{figure}

For the soft-on-hard gel ($n=0.5$), a critical wavelength of zero was obtained for all values of the diffusion coefficient as creasing was found to be the first mode of instability. This result was also found in the simulations with $n=0.1$ and $n=0.8$. For the hard-on-soft gels on the other hand ($n=2$ and $n=5$), wrinkling was found as the first mode of instability in most of the simulations, with a tendency of a reduction in the critical wavelength $\Lambda_c$ as the diffusion coefficient $D$ was reduced. For $n=2$ and $Nv_s=0.001$, the critical wavelength drops to zero when the diffusivity goes below $10^{-5}$ m$^2$/s, as creasing (rather than wrinkling) was found as the critical mode of instability in these simulations. This is illustrated in Figure \ref{f:buckling_profile} showing the instability mode for the fast and the slow diffusion processes for $Nv_s=0.001$ and $n=2$. It can be seen that for fast swelling, a structured wrinkling pattern is obtained where the size of the initial imperfection is negligible compared to the length scale of the wrinkles. For the slow swelling simulation, on the other hand, it can be noted that the out-of-plane deformation is limited to the upper part of the film and that a localized folding process has started at the surface of the gel. The location of the developing crease depends on the initial surface imperfection in the model. The results shown in Figures \ref{f:Effect_D_wl} and \ref{f:buckling_profile} were obtained with a random imperfection in the upper surface of the plate (i.e. Equation (\ref{eq:imper_rand})), however, similar results were obtained using the harmonic imperfection (i.e. Equation (\ref{eq:imper_sin})). In addition, the shift from wrinkling to creasing as the diffusion coefficient decrease from $10^{-5}$ m$^2$/s to $10^{-6}$ m$^2$/s was found to be insensitive of the mesh size (tested in the range between 20 and 50 elements over the film thickness). 

The wrinkling to creasing transition for a hard-on-soft gel clearly demonstrates how the time-dependent nature of swelling can cause a discrepancy between the buckling mode obtained in experiments compared to that predicted by an equilibrium analysis. Further, the fact that $Nv_s=0.001$ and $n=2$ reach instability by creasing before wrinkling at low values of $D$ explains the large difference in the critical swelling ratio between $Nv_s=0.001$ and $Nv_s=0.01$ during slow swelling for $n=2$ seen in Figure \ref{f:Effect_plane}. 

The authors hypothesize that creasing can occur during transient swelling of hard-on-soft gels with low film to substrate stiffness ratios due to a small difference between the strain energy in the creased and wrinkled configurations. For a sufficiently slow swelling process (i.e. a high ratio between the film thickness and the diffusion coefficient) in a sufficiently soft gel, instability will be triggered while the compressive stresses are focused in the upper part of the film, causing creasing to occur rather than wrinkling. However, further experimental and theoretical research is needed to fully understand the wrinkling to creasing transition during transient swelling for a hard-on-soft system and to quantify the critical conditions for this transition to occur. 

\subsection{Dependence on initial dimensions}\label{sec:res_inidim}
\subsubsection{Preliminaries}
For the length of the plate $L$, the generally good agreement between the LPA results, assuming an infinitely wide plate, and the FEM results, giving the plate a finite length, indicates that the ratio H/L is small enough for the results herein to be representative for infinitely wide plates. To discuss the influence on the obtained results by the chosen values for the film vs plate ratio $\eta$ and total plate thickness $H$, we will consider equilibrium and transient swelling separately. 

\subsubsection{Equilibrium swelling}
\begin{figure}[h!]
	\centering
    	\begin{subfigure}[t]{0.48\textwidth}
		\includegraphics[width=\textwidth]{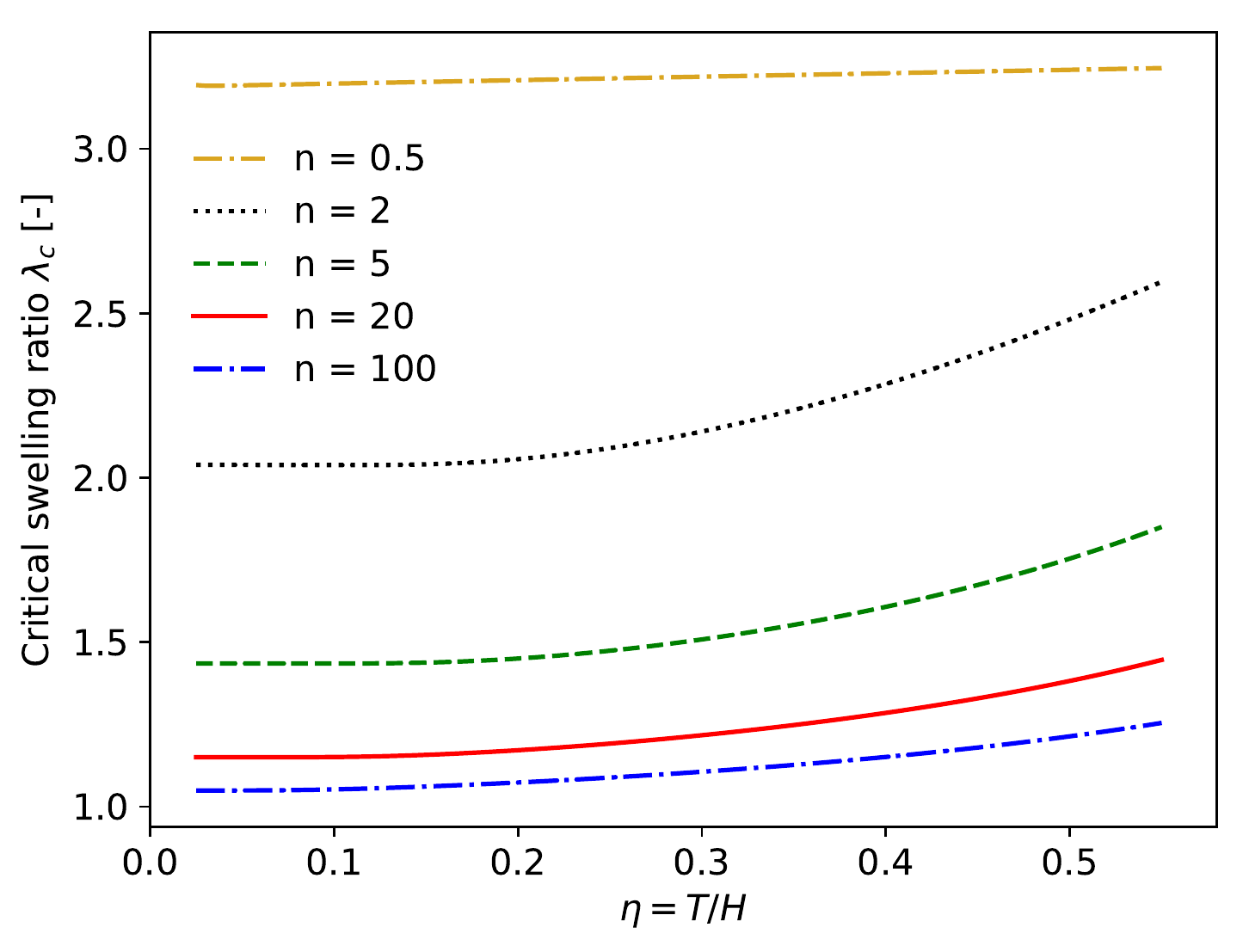}
		\caption{}
         \label{f:LPA_swelling}
	\end{subfigure}
	\begin{subfigure}[t]{0.48\textwidth}
		\includegraphics[width=\textwidth]{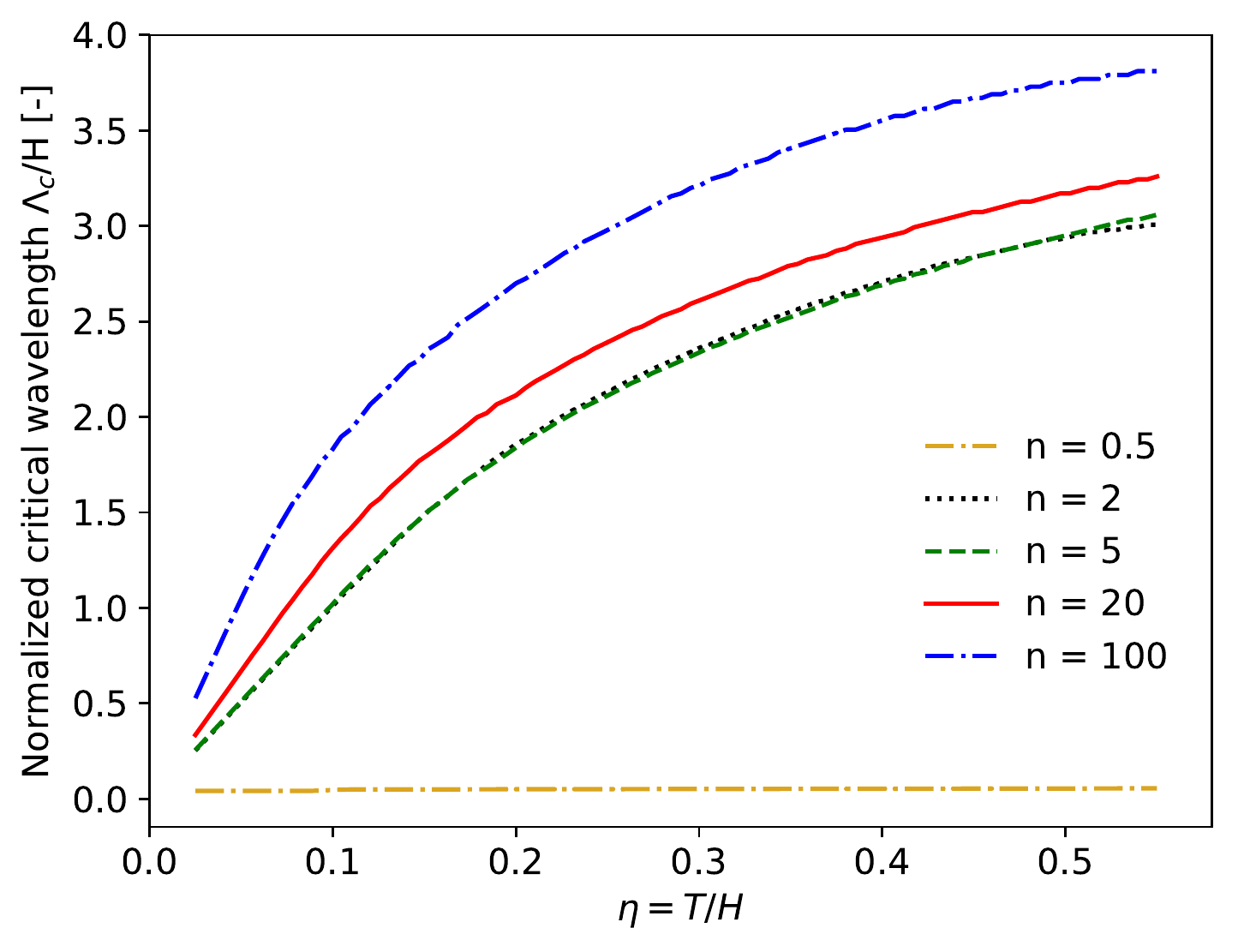}
		\caption{}
        \label{f:LPA_wave}
	\end{subfigure}
  \caption{Critical swelling ratio (a) and normalized critical wavelength (b) as a function of the film to total plate height ratio for equilibrium swelling. Results obtained using the LPA framework with $Nv_s=0.001$. The curves for $n=2$ and $n=5$ are nearly overlapping in (b).}
  \label{f:LPA_swellingwave}
\end{figure}

For a discussion on the geometrical dependencies under equilibrium swelling, the LPA framework is an effective method. First, looking at the film vs plate thickness ratio $\eta$, the dependence on the critical swelling ratio and the normalized critical wavelength $\bar{\Lambda}_c=\Lambda_c/H$ are shown in Figures \ref{f:LPA_swelling} and \ref{f:LPA_wave}, respectively. The data in the plots were obtained using $Nv_s=0.001$, although similar trends could be obtained with $Nv_s=0.01$. Starting with the soft-on-hard case ($n=0.5$), both the critical swelling ratio and the normalized critical wavelength can be seen to be nearly independent of $\eta$. For the hard-on-soft cases ($n>1$) on the other hand, we see that near the film to plate height ratio of $\eta=0.1$ used herein the critical swelling ratio (Figure \ref{f:LPA_swelling}) is relatively stable with respect to the film fraction. However, there is a clear tendency of an increasing critical swelling ratio as $\eta$ gets larger than 0.25, with a stronger dependence observed for lower stiffness ratios. The predicted normalized critical wavelength (Figure \ref{f:LPA_wave}) for the hard-on-soft cases is changing significantly near $\eta=0.1$, with increasing wavelengths as the film to plate thickness ratio is increased. 

Second, for a change in the initial height of the plate (keeping the ratio $\eta$ constant), both the critical swelling ratio and the normalized critical wavelength (as presented in Figure \ref{f:LPA_swellingwave}) were found to be independent of $H$ under the assumption of equilibrium swelling of an infinitely wide plate.   

\subsubsection{Transient swelling}
\begin{figure}[h!]
\centering
  \includegraphics[width=9cm]{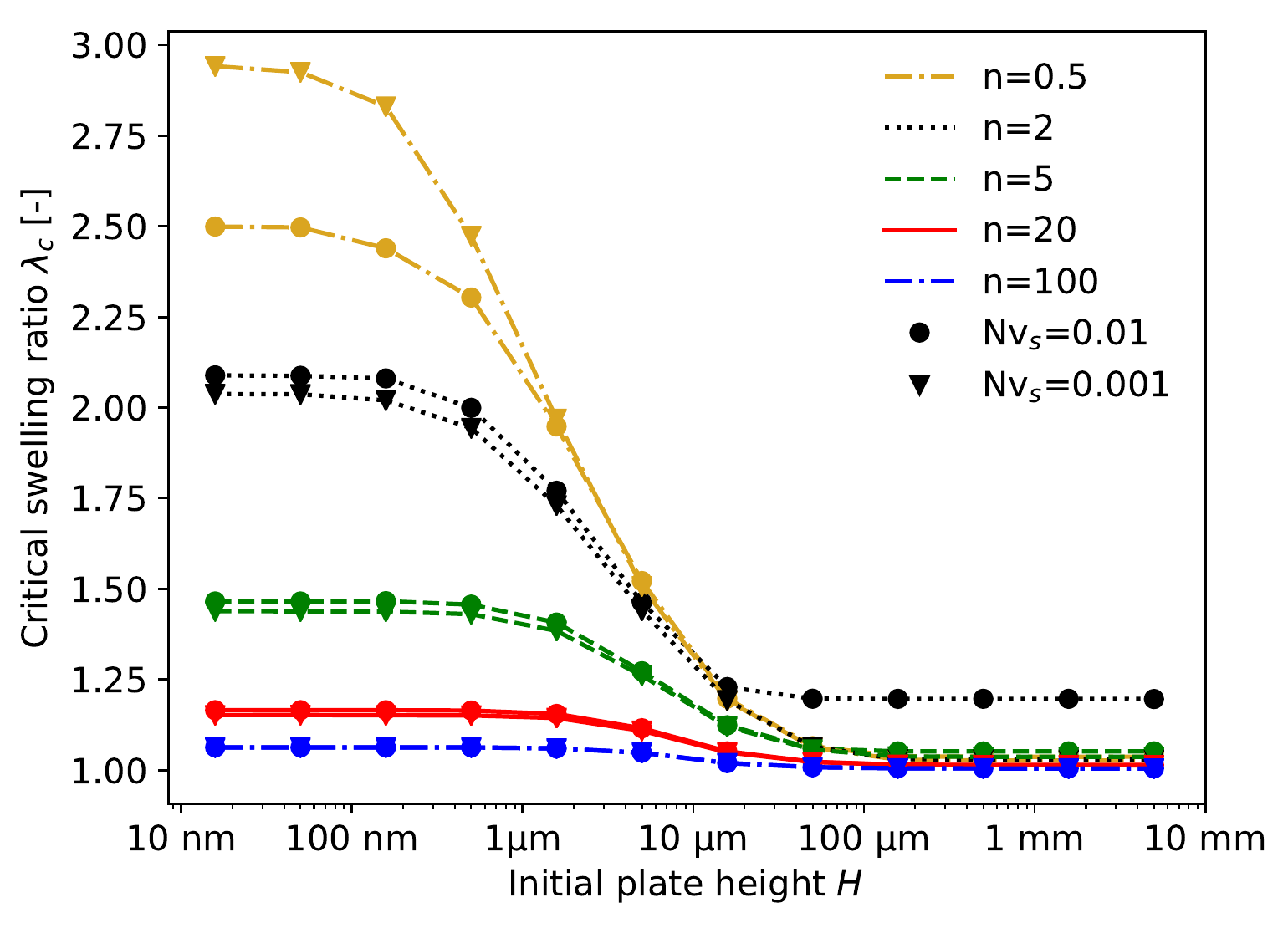}
  \caption{Effect of the varying the initial plate thickness (retaining the shape of the plate) on the critical swelling ratio for a constant diffusivity of $D=10^{-9}$ m$^2$/s.}
 \label{f:Effect_H_lamb}
\end{figure}

For a discussion on how the results for transient swelling depend on the initial thickness of the plate $H$ (keeping $\eta$ constant), it is worth noting that the defined problem introduces a relation between the length scale of the swelling plate and the diffusion coefficient of the gel material. This relation means that scaling $D$ with a factor $x^2$ would be equivalent to scaling the dimensions of the plate with a factor 1/$x$. Hence, the results presented in the previous sections for varying diffusivity and constant dimensions could have been obtained with a constant diffusivity and varying the plate dimensions. This scaling is demonstrated in Figure \ref{f:Effect_H_lamb}, plotting the critical swelling ratio as a function of the initial plate thickness for a constant diffusivity of $D=10^{-9}$ m$^2$/s.

It is important to note the difference in the dependence on the initial plate height $H$ between equilibrium and transient swelling. This difference highlights the importance of the length scale of a swelling hydrogel system for whether simple equilibrium analyses would yield precise predictions for the initiation of buckling, with improved precision of equilibrium estimates for thinner gels. In addition, the length scale dependence in transient swelling means that the regions I, II, and III and their respective diffusivity values as given in Section \ref{sec:stress_chem_profile} are specific for the plate dimensions used. 

A further discussion on how a change in the ratios $H/L$ and $T/H$ would influence the initiation of buckling during transient swelling is considered out-of-scope for the present work. 

\section{Conclusion} \label{sec:conc}
In this paper, we show that the onset of instability in hydrogels with gradient stiffness is highly influenced by the kinetic nature of the swelling process. Both the critical swelling ratio and the time to buckling initiation were found to increase as the diffusion coefficient of the material was reduced. In addition, these measures were found to depend on the absolute stiffness of the film and the substrate and not only the ratio between the two. 
 
For the geometrical configuration studied herein, changing the diffusivity within the physical region for hydrogels in water $\left(10^{-9}-10^{-11} \, \text{m}^2/\text{s}\right)$ would have a negligible effect on the swelling ratio at the onset of buckling, while the time to the onset of buckling could change significantly, with an increasing effect as the stiffness ratio between the stiff film and the soft substrate is reduced. 

For soft-on-hard gels, creasing was found as the first mode of instability independent of the diffusion coefficient. For hard-on-soft gels on the other hand, the diffusivity of the gel material could alter the buckling pattern by reducing the wavelength as the diffusivity was reduced. This is most evident for the combination of material parameters $Nv_s=0.001$ and $n=2$, where a stable wrinkling pattern was predicted by an equilibrium analysis, while creasing was triggered as the first mode of instability when using a physical value for the diffusion coefficient of the gel.

The results presented herein for a change in the diffusivity with constant plate dimensions could also be read as a change in plate dimensions (retaining the shape) with a constant value for the diffusivity. Whether simplified analyses based on an assumption of equilibrium swelling would be predictive for the onset of buckling in a layered hydrogel system would hence depend on the length scale of the problem. 

For further work, we suggest that the transition from wrinkling to creasing in hard-on-soft gels is studied both theoretically and experimentally. Further, viscoelastic effects of the polymer network should be included in the constitutive modeling to study the influence this will have on the initiation of buckling. In addition, analyses of the post-buckling behavior of confined swelling gels where the three-dimensional nature of the buckling pattern is accounted for could increase the understanding of the fundamental mechanisms governing swelling induced buckling and pattern evolution.

\section*{Conflicts of interest}
There are no conflicts to declare.

\section*{Acknowledgments}
This work was supported by the Norwegian Research Council (Project no 240299/F20). 
%%%END OF MAIN TEXT%%%
\appendix
\section{Gel diffusion and heat transfer analogy} \label{app:heat_ana}
This appendix outlines the analogy between gel swelling and heat transfer, paving the way for implementing coupled solvent diffusion and large deformations in Abaqus using default temperature-displacement elements. 

The energy balance in heat transfer as it is implemented in Abaqus reads
\begin{linenomath}
\begin{equation}
\int_V\rho c\frac{\partial T}{\partial t}dV +\int_S \mathbf{q}\cdot\mathbf{n}dS =  \int_V r dV
\label{eq:abaqus_energy}
\end{equation}
\end{linenomath}
where $\rho$ is the material density, $c$ is the specific heat capacity, $T$ is temperature, $\mathbf{q}$ is the heat flux flowing into the body, $\mathbf{n}$ the outward unit normal, and $r$ is the heat supplied internally into the body. $V$ and $S$ define the volume and the surface of the body in $\Omega$, respectively. The heat flux is given by
\begin{linenomath}
\begin{equation}
\mathbf{q}  =- k\frac{\partial T}{\partial \mathbf{x} }
\label{eq:abaqus_flux}
\end{equation}
\end{linenomath}
where $k$ is the thermal conductivity of the material. By assuming $C=C(\bar{\mu})$ the conservation of solvent molecules in a gel and its solution reads \cite{Chester2010,Toh2013}
\begin{linenomath}
\begin{equation}
\int_V\frac{1}{J}\frac{\partial C}{\partial \bar{\mu}}\frac{\partial \bar{\mu}}{\partial t}dV+\int_S\ \boldsymbol{\phi}\cdot \mathbf{n}dS=0
\label{eq:gel_conser}
\end{equation}
\end{linenomath}
while the flux of the solvent molecules in a gel was given in Equation (\ref{eq:jk}). By comparing Equation (\ref{eq:abaqus_energy}) to Equation (\ref{eq:gel_conser}) and Equation (\ref{eq:abaqus_flux}) to Equation (\ref{eq:jk}), we find that the equivalents of temperature, specific heat capacity, mass density, and conductivity can be identified as $\bar{\mu}$, $\frac{\partial C}{\partial \bar{\mu}}$, $\frac{1}{J}$, and $\frac{CD}{J}=\frac{J-1}{vJ}D$ respectively. These values are calculated as internal variables in the Fortran code using the USDFLD subroutine in Abaqus. The internal heat source in Abaqus, $r$, is set to zero. 

To find an analytical expression for $\frac{\partial C}{\partial \bar{\mu}}$ to implement as the specific heat capacity is non-trivial. A strategy suggested by Toh \textit{et al.} \cite{Toh2013} is to derive an expression for $\bar{\mu}$ from Equation (\ref{eq:cauchy_stress}), using Equation (\ref{eq:mol_incom}) to relate $J$ and $C$
\begin{linenomath}
\begin{equation}
\bar{\mu}=\ln\left(\frac{vC}{1+vC}\right)+\frac{1}{1+vC}+\frac{\chi}{\left(1+vC\right)^2}+Nv\frac{\bar{I}_1\left(1+vC\right)^{2/3}-3}{3\left(1+vC\right)} -\frac{\text{tr} \bar{\boldsymbol{\sigma}}}{3}
\label{eq:mu_bar}
\end{equation}
\end{linenomath}
where $\text{tr}\bar{\boldsymbol{\sigma}}$ denotes the trace of the normalized Cauchy stress tensor and $\bar{I}_1=\left(1+vC\right)^{-2/3}I_1$. Assuming $C=C(\bar{\mu})$, the right-hand side of Equation (\ref{eq:mu_bar}) is differentiated with respect to $\bar{\mu}$ by use of symbolic differentiation. Using $\partial \bar{\mu}/\partial \bar{\mu}=1$, an expression for $\frac{\partial C}{\partial \bar{\mu} }$ is obtained as
\begin{linenomath}
\begin{equation}
\frac{\partial C}{\partial \bar{\mu} }= - \frac{9J^{10/3}\left(J-1\right)}{v\left(\left(9Nv+18\chi-9\right)J^{4/3}-9NvJ^{7/3}-18\chi J^{1/3}+\bar{I}_1Nv\left(J-1\right)J^2\right)}
\label{eq:diffCmu}
\end{equation}
\end{linenomath}
Note that Equation (\ref{eq:diffCmu}) is slightly different from the expression given in Toh \textit{et al.} \cite{Toh2013}. The good benchmark results shown in Figure \ref{f:validate} indicates that Equation (\ref{eq:diffCmu}) yields good accuracy for the problem studied herein.

%%%REFERENCES%%%
\section*{References}
\bibliography{cite} 

\begin{thebibliography}{10}
\expandafter\ifx\csname url\endcsname\relax
  \def\url#1{\texttt{#1}}\fi
\expandafter\ifx\csname urlprefix\endcsname\relax\def\urlprefix{URL }\fi
\expandafter\ifx\csname href\endcsname\relax
  \def\href#1#2{#2} \def\path#1{#1}\fi

\bibitem{Beebe2000}
D.~J. Beebe, J.~S. Moore, J.~M. Bauer, Q.~Yu, R.~H. Liu, C.~Devadoss, B.-H. Jo,
  {Functional hydrogel structures for autonomous flow control inside
  microfluidic channels}, Nature 404 (2000) 588--590.
\newblock \href {http://dx.doi.org/10.1038/35007047}
  {\path{doi:10.1038/35007047}}.

\bibitem{Chan2008}
G.~Chan, D.~J. Mooney, {New materials for tissue engineering: towards greater
  control over the biological response}, Trends in Biotechnology 26 (2008)
  382--392.
\newblock \href {http://dx.doi.org/10.1016/j.tibtech.2008.03.011}
  {\path{doi:10.1016/j.tibtech.2008.03.011}}.

\bibitem{Annabi2014}
N.~Annabi, A.~Tamayol, J.~A. Uquillas, M.~Akbari, L.~E. Bertassoni, C.~Cha,
  G.~Camci-Unal, M.~R. Dokmeci, N.~A. Peppas, A.~Khademhosseini, {25th
  Anniversary Article: Rational Design and Applications of Hydrogels in
  Regenerative Medicine}, Advanced materials (Deerfield Beach, Fla.) 26 (2014)
  85--124.
\newblock \href {http://dx.doi.org/10.1002/adma.201303233}
  {\path{doi:10.1002/adma.201303233}}.

\bibitem{Seliktar2012}
D.~Seliktar, {Designing cell-compatible hydrogels for biomedical applications},
  Science 336 (2012) 1124--1128.
\newblock \href {http://dx.doi.org/10.1126/science.1214804}
  {\path{doi:10.1126/science.1214804}}.

\bibitem{Bysell2006}
H.~Bysell, M.~Malmsten, {Visualizing the interaction between poly-L-lysine and
  poly(acrylic acid) microgels using microscopy techniques: Effect of
  electrostatics and peptide size}, Langmuir 22 (2006) 5476--5484.
\newblock \href {http://dx.doi.org/10.1021/la060452a}
  {\path{doi:10.1021/la060452a}}.

\bibitem{Li2016a}
J.~Li, D.~J. Mooney, {Designing hydrogels for controlled drug delivery}, Nature
  Reviews Materials 1 (2016) 16071.

\bibitem{Culver2017}
H.~R. Culver, J.~R. Clegg, N.~A. Peppas, {Analyte-Responsive Hydrogels:
  Intelligent Materials for Biosensing and Drug Delivery}, Accounts of Chemical
  Research 50 (2017) 170--178.
\newblock \href {http://dx.doi.org/10.1021/acs.accounts.6b00533}
  {\path{doi:10.1021/acs.accounts.6b00533}}.

\bibitem{Tierney2008}
S.~Tierney, D.~R. Hjelme, B.~T. Stokke, {Determination of swelling of
  responsive gels with nanometer resolution. Fiber-optic based platform for
  hydrogels as signal transducers}, Analytical Chemistry 80 (2008) 5086--5093.
\newblock \href {http://dx.doi.org/10.1021/ac800292k}
  {\path{doi:10.1021/ac800292k}}.

\bibitem{Buenger2012}
D.~Buenger, F.~Topuz, J.~Groll, {Hydrogels in sensing applications}, Progress
  in Polymer Science 37 (2012) 1678--1719.
\newblock \href {http://dx.doi.org/10.1016/j.progpolymsci.2012.09.001}
  {\path{doi:10.1016/j.progpolymsci.2012.09.001}}.

\bibitem{Tanaka1987}
T.~Tanaka, S.-T. Sun, Y.~Hirokawa, S.~Katayama, J.~Kucera, Y.~Hirose, T.~Amiya,
  {Mechanical instability of gels at the phase transition}, Nature 325 (1987)
  796--798.
\newblock \href {http://dx.doi.org/10.1038/325796a0}
  {\path{doi:10.1038/325796a0}}.

\bibitem{Chan2006}
E.~P. Chan, A.~J. Crosby, {Spontaneous formation of stable aligned wrinkling
  patterns}, Soft Matter 2 (2006) 324--328.
\newblock \href {http://dx.doi.org/10.1039/b515628a}
  {\path{doi:10.1039/b515628a}}.

\bibitem{Trujillo2008}
V.~Trujillo, J.~Kim, R.~C. Hayward, {Creasing instability of surface-attached
  hydrogels}, Soft Matter 4 (2008) 564.
\newblock \href {http://dx.doi.org/10.1039/b713263h}
  {\path{doi:10.1039/b713263h}}.

\bibitem{Chan2008a}
E.~P. Chan, E.~J. Smith, R.~C. Hayward, A.~J. Crosby, {Surface wrinkles for
  smart adhesion}, Advanced Materials 20 (2008) 711--716.
\newblock \href {http://dx.doi.org/10.1002/adma.200701530}
  {\path{doi:10.1002/adma.200701530}}.

\bibitem{Guvendiren2009}
M.~Guvendiren, S.~Yang, J.~A. Burdick, {Swelling-Induced surface patterns in
  hydrogels with gradient crosslinking density}, Advanced Functional Materials
  19 (2009) 3038--3045.
\newblock \href {http://dx.doi.org/10.1002/adfm.200900622}
  {\path{doi:10.1002/adfm.200900622}}.

\bibitem{Breid2009}
D.~Breid, A.~J. Crosby, {Surface wrinkling behavior of finite circular plates},
  Soft Matter 5 (2009) 425--431.
\newblock \href {http://dx.doi.org/10.1039/B807820C}
  {\path{doi:10.1039/B807820C}}.

\bibitem{Chung2009}
J.~Y. Chung, A.~J. Nolte, C.~M. Stafford, {Diffusion-controlled, self-organized
  growth of symmetric wrinkling patterns}, Advanced Materials 21 (2009)
  1358--1362.
\newblock \href {http://dx.doi.org/10.1002/adma.200803209}
  {\path{doi:10.1002/adma.200803209}}.

\bibitem{Yang2010}
S.~Yang, K.~Khare, P.~C. Lin, {Harnessing surface wrinkle patterns in soft
  matter}, Advanced Functional Materials 20 (2010) 2550--2564.
\newblock \href {http://dx.doi.org/10.1002/adfm.201000034}
  {\path{doi:10.1002/adfm.201000034}}.

\bibitem{Li2012a}
B.~Li, Y.-P. Cao, X.-Q. Feng, H.~Gao, {Mechanics of morphological instabilities
  and surface wrinkling in soft materials: a review}, Soft Matter 8 (2012)
  5728--5745.
\newblock \href {http://dx.doi.org/10.1039/c2sm00011c}
  {\path{doi:10.1039/c2sm00011c}}.

\bibitem{Chen2014}
D.~Chen, J.~Yoon, D.~Chandra, A.~J. Crosby, R.~C. Hayward, {Stimuli-responsive
  buckling mechanics of polymer films}, Journal of Polymer Science, Part B:
  Polymer Physics 52 (2014) 1441--1461.
\newblock \href {http://dx.doi.org/10.1002/polb.23590}
  {\path{doi:10.1002/polb.23590}}.

\bibitem{Zhou2017}
Z.~Zhou, Y.~Li, W.~Wong, T.~Guo, S.~Tang, J.~Luo, {Transition of
  surface–interface creasing in bilayer hydrogels}, Soft Matter 13 (2017)
  6011--6020.
\newblock \href {http://dx.doi.org/10.1039/C7SM01013C}
  {\path{doi:10.1039/C7SM01013C}}.

\bibitem{Guvendiren2010}
M.~Guvendiren, J.~A. Burdick, S.~Yang, {Kinetic study of swelling-induced
  surface pattern formation and ordering in hydrogel films with depth-wise
  crosslinking gradient}, Soft Matter 6 (2010) 2044--2049.
\newblock \href {http://dx.doi.org/10.1039/b927374c}
  {\path{doi:10.1039/b927374c}}.

\bibitem{Guvendiren2010a}
M.~Guvendiren, J.~A. Burdick, S.~Yang, {Solvent induced transition from
  wrinkles to creases in thin film gels with depth-wise crosslinking
  gradients}, Soft Matter 6 (2010) 5795--5801.
\newblock \href {http://dx.doi.org/10.1039/c0sm00317d}
  {\path{doi:10.1039/c0sm00317d}}.

\bibitem{Sultan2008}
E.~Sultan, A.~Boudaoud, {The Buckling of a Swollen Thin Gel Layer Bound to a
  Compliant Substrate}, Journal of Applied Mechanics 75 (2008) 51002--51005.

\bibitem{Prot2013}
V.~Prot, H.~M. Sveinsson, K.~Gawel, M.~Gao, B.~Skallerud, B.~T. Stokke,
  {Swelling of a hemi-ellipsoidal ionic hydrogel for determination of material
  properties of deposited thin polymer films: an inverse finite element
  approach}, Soft Matter 9 (2013) 5815--5827.
\newblock \href {http://dx.doi.org/10.1039/c3sm50805f}
  {\path{doi:10.1039/c3sm50805f}}.

\bibitem{Sherstova2016}
T.~Sherstova, B.~T. Stokke, B.~Skallerud, G.~Maurstad, V.~E. Prot,
  {Nanoindentation and finite element modelling of chitosan–alginate
  multilayer coated hydrogels}, Soft Matter 12 (2016) 7338--7349.
\newblock \href {http://dx.doi.org/10.1039/C6SM00827E}
  {\path{doi:10.1039/C6SM00827E}}.

\bibitem{Liu2012}
Z.~Liu, S.~Swaddiwudhipong, W.~Hong, {Pattern formation in plants via
  instability theory of hydrogels}, Soft Matter 9 (2012) 577--587.
\newblock \href {http://dx.doi.org/10.1039/C2SM26642C}
  {\path{doi:10.1039/C2SM26642C}}.

\bibitem{Limbert2018}
G.~Limbert, E.~Kuhl, {On skin microrelief and the emergence of expression
  micro-wrinkles}, Soft Matter 14 (2018) 1292--1300.
\newblock \href {http://dx.doi.org/10.1039/C7SM01969F}
  {\path{doi:10.1039/C7SM01969F}}.

\bibitem{Bowden1998}
N.~Bowden, S.~Brittain, A.~G. Evans, J.~W. Hutchinson, G.~M. Whitesides,
  {Spontaneous formation of ordered structures in thin films of metals
  supported on an elastomeric polymer}, Nature 393 (1998) 146--149.
\newblock \href {http://dx.doi.org/10.1038/30193} {\path{doi:10.1038/30193}}.

\bibitem{Groenewold2001}
J.~Groenewold, {Wrinkling of plates coupled with soft elastic media}, Physica
  A: Statistical Mechanics and its Applications 298 (2001) 32--45.
\newblock \href {http://dx.doi.org/10.1016/S0378-4371(01)00209-6}
  {\path{doi:10.1016/S0378-4371(01)00209-6}}.

\bibitem{Yin2018}
S.~F. Yin, B.~Li, Y.~P. Cao, X.~Q. Feng, {Surface wrinkling of anisotropic
  films bonded on a compliant substrate}, International Journal of Solids and
  Structures 141-142 (2018) 219--231.
\newblock \href {http://dx.doi.org/10.1016/j.ijsolstr.2018.02.024}
  {\path{doi:10.1016/j.ijsolstr.2018.02.024}}.

\bibitem{Stafford2004}
C.~M. Stafford, C.~Harrison, K.~L. Beers, A.~Karim, E.~J. Amis, M.~R.
  VanLandingham, H.~C. Kim, W.~Volksen, R.~D. Miller, E.~E. Simonyi, {A
  buckling-based metrology for measuring the elastic moduli of polymeric thin
  films}, Nature Materials 3 (2004) 545--550.
\newblock \href {http://dx.doi.org/10.1038/nmat1175}
  {\path{doi:10.1038/nmat1175}}.

\bibitem{Wu2013}
Z.~Wu, N.~Bouklas, R.~Huang, {Swell-induced surface instability of hydrogel
  layers with material properties varying in thickness direction},
  International Journal of Solids and Structures 50 (2013) 578--587.
\newblock \href {http://dx.doi.org/10.1016/j.ijsolstr.2012.10.022}
  {\path{doi:10.1016/j.ijsolstr.2012.10.022}}.

\bibitem{Wu2017}
Z.~Wu, N.~Bouklas, Y.~Liu, R.~Huang, {Onset of swell-induced surface
  instability of hydrogel layers with depth-wise graded material properties},
  Mechanics of Materials 105 (2017) 138--147.
\newblock \href {http://dx.doi.org/10.1016/j.mechmat.2016.11.005}
  {\path{doi:10.1016/j.mechmat.2016.11.005}}.

\bibitem{Hong2009}
W.~Hong, Z.~Liu, Z.~Suo, {Inhomogeneous swelling of a gel in equilibrium with a
  solvent and mechanical load}, International Journal of Solids and Structures
  46 (2009) 3282--3289.
\newblock \href {http://dx.doi.org/10.1016/j.ijsolstr.2009.04.022}
  {\path{doi:10.1016/j.ijsolstr.2009.04.022}}.

\bibitem{Kang2010a}
M.~K. Kang, R.~Huang, {A Variational Approach and Finite Element Implementation
  for Swelling of Polymeric Hydrogels Under Geometric Constraints}, Journal of
  Applied Mechanics 77 (2010) 61004.
\newblock \href {http://dx.doi.org/10.1115/1.4001715}
  {\path{doi:10.1115/1.4001715}}.

\bibitem{Marcombe2010}
R.~Marcombe, S.~Cai, W.~Hong, X.~Zhao, Y.~Lapusta, Z.~Suo, {A theory of
  constrained swelling of a pH-sensitive hydrogel}, Soft Matter 6 (2010)
  784--793.
\newblock \href {http://dx.doi.org/10.1039/b917211d}
  {\path{doi:10.1039/b917211d}}.

\bibitem{Toh2013}
W.~Toh, Z.~Liu, T.~Y. Ng, W.~Hong, {Inhomogeneous Large Deformation Kinetics of
  Polymeric Gels}, International Journal of Applied Mechanics 05 (2013)
  1350001.
\newblock \href {http://dx.doi.org/10.1142/S1758825113500014}
  {\path{doi:10.1142/S1758825113500014}}.

\bibitem{Duan2013}
Z.~Duan, J.~Zhang, Y.~An, H.~Jiang, {Simulation of the Transient Behavior of
  Gels Based on an Analogy Between Diffusion and Heat Transfer}, Journal of
  Applied Mechanics 80 (2013) 41017.
\newblock \href {http://dx.doi.org/10.1115/1.4007789}
  {\path{doi:10.1115/1.4007789}}.

\bibitem{Zhang2009}
J.~Zhang, X.~Zhao, Z.~Suo, H.~Jiang, {A finite element method for transient
  analysis of concurrent large deformation and mass transport in gels}, Journal
  of Applied Physics 105 (2009) 093522.
\newblock \href {http://dx.doi.org/10.1063/1.3106628}
  {\path{doi:10.1063/1.3106628}}.

\bibitem{Chester2015}
S.~A. Chester, C.~V. {Di Leo}, L.~Anand, {A finite element implementation of a
  coupled diffusion-deformation theory for elastomeric gels}, International
  Journal of Solids and Structures 52 (2015) 1--18.
\newblock \href {http://dx.doi.org/10.1016/j.ijsolstr.2014.08.015}
  {\path{doi:10.1016/j.ijsolstr.2014.08.015}}.

\bibitem{Bouklas2015}
N.~Bouklas, C.~M. Landis, R.~Huang, {A nonlinear, transient finite element
  method for coupled solvent diffusion and large deformation of hydrogels},
  Journal of the Mechanics and Physics of Solids 79 (2015) 21--43.
\newblock \href {http://dx.doi.org/10.1016/j.jmps.2015.03.004}
  {\path{doi:10.1016/j.jmps.2015.03.004}}.

\bibitem{Kang2010}
M.~K. Kang, R.~Huang, {Swell-induced surface instability of confined hydrogel
  layers on substrates}, Journal of the Mechanics and Physics of Solids 58
  (2010) 1582--1598.
\newblock \href {http://dx.doi.org/10.1016/j.jmps.2010.07.008}
  {\path{doi:10.1016/j.jmps.2010.07.008}}.

\bibitem{Weiss2013}
F.~Weiss, S.~Cai, Y.~Hu, M.~{Kyoo Kang}, R.~Huang, Z.~Suo, {Creases and
  wrinkles on the surface of a swollen gel}, Journal of Applied Physics 114
  (2013) 073507.
\newblock \href {http://dx.doi.org/10.1063/1.4818943}
  {\path{doi:10.1063/1.4818943}}.

\bibitem{Budday2015}
S.~Budday, E.~Kuhl, J.~W. Hutchinson, {Period-doubling and period-tripling in
  growing bilayered systems}, Philosophical Magazine 95 (2015) 3208--3224.
\newblock \href {http://dx.doi.org/10.1080/14786435.2015.1014443}
  {\path{doi:10.1080/14786435.2015.1014443}}.

\bibitem{Xu2014a}
F.~Xu, M.~Potier-Ferry, S.~Belouettar, Y.~Cong, {3D finite element modeling for
  instabilities in thin films on soft substrates}, International Journal of
  Solids and Structures 51 (2014) 3619--3632.
\newblock \href {http://dx.doi.org/10.1016/j.ijsolstr.2014.06.023}
  {\path{doi:10.1016/j.ijsolstr.2014.06.023}}.

\bibitem{Xu2015}
F.~Xu, Y.~Koutsawa, M.~Potier-Ferry, S.~Belouettar, {Instabilities in thin
  films on hyperelastic substrates by 3D finite elements}, International
  Journal of Solids and Structures 69-70 (2015) 71--85.
\newblock \href {http://dx.doi.org/10.1016/j.ijsolstr.2015.06.007}
  {\path{doi:10.1016/j.ijsolstr.2015.06.007}}.

\bibitem{Xu2016}
F.~Xu, M.~Potier-Ferry, {A multi-scale modeling framework for instabilities of
  film/substrate systems}, Journal of the Mechanics and Physics of Solids 86
  (2016) 150--172.
\newblock \href {http://dx.doi.org/10.1016/j.jmps.2015.10.003}
  {\path{doi:10.1016/j.jmps.2015.10.003}}.

\bibitem{Toh2016}
W.~Toh, Z.~Ding, T.~{Yong Ng}, Z.~Liu, {Wrinkling of a Polymeric Gel During
  Transient Swelling}, Journal of Applied Mechanics 82 (2015) 061004.
\newblock \href {http://dx.doi.org/10.1115/1.4030327}
  {\path{doi:10.1115/1.4030327}}.

\bibitem{Yu2018}
C.~Yu, K.~Malakpoor, J.~M. Huyghe, {A three-dimensional transient mixed hybrid
  finite element model for superabsorbent polymers with strain-dependent
  permeability}, Soft Matter 14 (2018) 3834--3848.
\newblock \href {http://dx.doi.org/10.1039/C7SM01587A}
  {\path{doi:10.1039/C7SM01587A}}.

\bibitem{Dortdivanlioglu2019}
B.~Dortdivanlioglu, C.~Linder, {Diffusion-driven swelling-induced instabilities
  of hydrogels}, Journal of the Mechanics and Physics of Solids 125 (2019)
  38--52.
\newblock \href {http://dx.doi.org/10.1016/j.jmps.2018.12.010}
  {\path{doi:10.1016/j.jmps.2018.12.010}}.

\bibitem{Caccavo2018}
D.~Caccavo, S.~Cascone, G.~Lamberti, A.~A. Barba, {Hydrogels: experimental
  characterization and mathematical modelling of their mechanical and diffusive
  behaviour}, Chem. Soc. Rev. 47 (2018) 2357--2373.
\newblock \href {http://dx.doi.org/10.1039/C7CS00638A}
  {\path{doi:10.1039/C7CS00638A}}.

\bibitem{Hong2008}
W.~Hong, X.~Zhao, J.~Zhou, Z.~Suo, {A theory of coupled diffusion and large
  deformation in polymeric gels}, Journal of the Mechanics and Physics of
  Solids 56 (2008) 1779--1793.
\newblock \href {http://dx.doi.org/10.1016/j.jmps.2007.11.010}
  {\path{doi:10.1016/j.jmps.2007.11.010}}.

\bibitem{flory1953}
P.~J. Flory, {Principles of polymer chemistry}, Cornell University Press, 1953.

\bibitem{Huggins1941}
M.~L. Huggins, {Solutions of Long Chain Compounds}, The Journal of Chemical
  Physics 9 (1941) 440--440.
\newblock \href {http://dx.doi.org/10.1063/1.1750930}
  {\path{doi:10.1063/1.1750930}}.

\bibitem{Flory1942}
P.~J. Flory, {Thermodynamics of High Polymer Solutions}, The Journal of
  Chemical Physics 10 (1942) 51--61.
\newblock \href {http://dx.doi.org/10.1063/1.1723621}
  {\path{doi:10.1063/1.1723621}}.

\bibitem{Feynman1963}
R.~P. Feynman, R.~B. Leighton, M.~L. Sands, {The Feynman lectures on physics},
  Addison-Wesley Pub. Co., 1963.

\bibitem{Abaqus2014}
Abaqus, 6.14-4, Dassault Syst{\`{e}}mes, 2014.

\bibitem{Ilseng}
A.~Ilseng, V.~Prot, {User subroutine for modeling transient swelling of
  hydrogels with Abaqus}.
\newblock \href {http://dx.doi.org/10.17632/yx3bj2rw3g.2}
  {\path{doi:10.17632/yx3bj2rw3g.2}}.

\bibitem{Cai2011a}
S.~Cai, D.~Breid, A.~J. Crosby, Z.~Suo, J.~W. Hutchinson, {Periodic patterns
  and energy states of buckled films on compliant substrates}, Journal of the
  Mechanics and Physics of Solids 59 (2011) 1094--1114.
\newblock \href {http://dx.doi.org/10.1016/j.jmps.2011.02.001}
  {\path{doi:10.1016/j.jmps.2011.02.001}}.

\bibitem{Breid2011}
D.~Breid, A.~J. Crosby, {Effect of stress state on wrinkle morphology}, Soft
  Matter 7 (2011) 4490--4496.
\newblock \href {http://dx.doi.org/10.1039/c1sm05152k}
  {\path{doi:10.1039/c1sm05152k}}.

\bibitem{bergheau2013finite}
J.~M. Bergheau, R.~Fortunier, {Finite Element Simulation of Heat Transfer},
  Wiley, 2013.

\bibitem{Dortdivanlioglu2018}
B.~Dortdivanlioglu, L.~B. Veiga, C.~Linder, {Mixed isogeometric analysis of
  strongly coupled diffusion in porous materials}, International Journal for
  Numerical Methods in Engineering 114 (2018) 28--46.
\newblock \href {http://dx.doi.org/10.1002/nme.5731}
  {\path{doi:10.1002/nme.5731}}.

\bibitem{Chester2010}
S.~A. Chester, L.~Anand, {A coupled theory of fluid permeation and large
  deformations for elastomeric materials}, Journal of the Mechanics and Physics
  of Solids 58 (2010) 1879--1906.
\newblock \href {http://dx.doi.org/10.1016/j.jmps.2010.07.020}
  {\path{doi:10.1016/j.jmps.2010.07.020}}.

\end{thebibliography}

\end{document}